\documentclass[a4paper]{article}

\usepackage{amsmath,amssymb}
\usepackage{color}
\usepackage{graphicx}
\usepackage{afterpage}
\definecolor{refkey}{gray}{.75}
\definecolor{labelkey}{gray}{.75}
\usepackage{subfigure}
\usepackage[normalem]{ulem}

\newcommand{\eps}{\varepsilon}

\newcommand{\tit}{{\tilde t}}
\newcommand{\tix}{{\tilde x}}
\newcommand{\tis}{{\tilde s}}
\newcommand{\timu}{{\tilde \mu}}
\newcommand{\tiq}{{\tilde q}}

\newcommand{\tih}{{\tilde h}}

\newcommand{\mfnd}{\phi} 

\newcommand{\lcut}{L_\infty}

\newcommand{\todo}[1]{}

\newcommand{\com}[1]{}

\newcounter{todo}
\renewcommand{\todo}[1]{\addtocounter{todo}{1}{\textcolor{black}{\\[0.5ex] \centerline{\fbox{\begin{minipage}{0.85\textwidth}%
\textbf{[\#\arabic{todo}]} \textit{\small #1}\end{minipage}}}}\\[0.5ex]}}

\renewcommand{\com}[1]{\footnote{\textbf{[Comment or question:} #1\textbf{]}}}

\begin{document}

\title{Controlled topological transitions in thin film phase separation}

\author{Matthew G.~Hennessy\footnotemark[2] \and
        Victor M.~Burlakov\footnotemark[2] \and 
	Alain Goriely\footnotemark[2] \and
	Andreas M\"unch\footnotemark[2] \and
	Barbara Wagner\footnotemark[3]\ \footnotemark[4]}

\date{Submitted 23/12/2013}
	
\maketitle
\pagestyle{myheadings}
\thispagestyle{plain}
\markboth{Hennessy et al.}{Controlled topological transitions}

\renewcommand{\thefootnote}{\fnsymbol{footnote}}
\footnotetext[2]{Mathematical Institute, University of Oxford, Andrew Wiles Building, Radcliffe Observatory Quarter, Woodstock Road, Oxford, OX2 6GG, UK}
\footnotetext[3]{Technische Universit\"at Berlin, Institute of Mathematics, Stra{\ss}e des 17. Juni 136, 10623 Berlin, Germany}
\footnotetext[4]{Weierstrass Institute, Mohrenstra{\ss}e 39, 10117 Berlin, Germany}

\begin{abstract}
In this paper the evolution of a binary mixture in a thin-film geometry with a wall at the top and bottom is considered. 
By bringing the mixture into its miscibility gap
so that no spinodal decomposition occurs in the bulk, a slight energetic bias of the walls towards each one of the constituents ensures the nucleation of thin boundary layers that grow until the constituents have moved into one of the two layers. These layers are separated by an interfacial region where the composition changes rapidly. Conditions that ensure the separation into two layers with a thin interfacial region are investigated based on a phase-field model. Using matched asymptotic expansions a corresponding sharp-interface problem for the location of the interface is established. 

It is then argued that this newly created two-layer system is not at its energetic minimum but destabilizes into a controlled self-replicating pattern of trapezoidal vertical stripes by minimizing the interfacial energy between the phases while conserving their area. 
A quantitative analysis of this mechanism is carried out via a thin-film model for the free interfaces, which is derived asymptotically from the sharp-interface model.


\end{abstract}



\section{Introduction}\label{sec:intro}

Structure formation in mixtures such as polymer blends and metal or semiconductor alloys is abundant in nature and in many technological processes. This phenomenon generally occurs when, due to a change in the external conditions such as temperature, pressure, applied stresses or a change in the composition, it becomes energetically preferable for the materials to be in a non-homogeneous structured state rather than the homogeneous state. The system will tend to a new minimum of its associated total free energy by undergoing phase transformations through spinodal decomposition or nucleation events, thereby forming new spatial domains with different composition, polymer phases, crystal structure or orientation, giving rise to new material properties. The new interfaces that bound these domains coarsen on much slower time scales until a global energetic minimum has been reached. For a review on structure forming processes in materials see, for example, \cite{Binder2001a}.

Throughout the whole process different physical effects and material properties such as interfacial stresses, elastic strains, chemical reactions at interfaces, electrostatic forces, bulk and interfacial diffusion may have to be taken into account. Moreover, in confined geometries, as it is the case for most nano-technological applications, the influence of nearby walls will have a significant impact on the phase separation by either introducing additional geometric length scales or through their surface energies which determine their wetting properties. 

For a thermally-quenched mixture in a confined geometry it has been shown,  experimentally \cite{Jones1991,Krausch1995} and theoretically \cite{Binder2011}, that phase separation can be induced by the interface energy of the nearby walls.    Various scenarios of these so-called surface-directed phase separation phenomena have been investigated on the basis of appropriate boundary value problems for the stochastic Cahn-Hilliard model including off-critical quenches, i.e., where no phase separation would occur for unconfined case, see e.g \cite{Puri1994}. Their numerical results exhibit cases where typical bulk phase separation occurs together with a wetting layer as well as cases where only the growing wetting layer emerges.  These studies were extended further using different free energies and different intermolecular potentials, see \cite{Brown1993,Yan2008} for a discussion of these models.

Similarly, during spin-coating of a mixture of two polymers blended in a common volatile solvent a stratified film of nano- to micrometer thickness that exhibits an internal interfacial microstructure is produced. For such processes it has been suggested in \cite{HJ05,JHSJ05,SWBS03,JBBMR08}, using PFB/F8BT  and PS/PMMA systems, that phase separation starts with the formation of a vertically stratified bilayer, followed by a destabilisation of the polymer-polymer interface and it is speculated that this is due to a solvent-concentration
gradient through the film. 

Developing a systematic quantitative understanding of such complex evolutionary processes is the key to predict and control the structure morphology and hence the material properties, such as the optical and electrical properties of the active component of organic polymer-polymer solar cells, or other advanced multifunctional materials.
In \cite{Hennessy2013} a new mechanism that induces a well-defined sequence of repeating structures in a geometrically-confined binary mixture is presented. A qualitative argument is given to explain how a horizontal bilayer state may transition into a striped state of alternating phases. The metastable horizontal layered state enters a cascade of rupture events that lead to a state with regular well-defined trapezoidal stripes, minimising their interfacial energies.

The focus of the present study is to develop the theory behind the transition
mechanism described in \cite{Hennessy2013}, and to
determine under which conditions the horizontal bilayer state can form while allowing for subsequent stripe formation. 
Classical ways to create a bilayer include 
imposing an external field that is switched off, placing the two layers on top of each other, or using initial compositional gradients
 that can selectively drive the coarsening of the structured state created by the initial phase separation and give rise to long-lived metastable bilayer states \cite{Jaiswal2012}. However, here we will
explore another possibility. In particular, we use surfaces that are biased towards one of the components and then bring the system slowly into the miscibility gap. The bias of the walls will create a slight compositional gradient across the thin film between the substrates that continues to build up as the species are driven towards separation.

Within the framework of a phase-field theory of Cahn--Hilliard type with appropriate surface energies at the walls, which we introduce in Section 2, we address the question of when the mixture phase separates into two horizontal layers with a diffuse interface that is thin compared to the transversal length scale of the thin film in Section 3.
The scale separation between the large homogeneous regions and the thin regions of steep compositional changes are then exploited in Section 4 to reduce the model to a sharp-interface model via the method of matched asymptotic expansions. 
Here we point out that for cases where the interfaces do not intersect an exterior boundary, such as a wall, the derivation of such models go back to Pego \cite{pego_front_1989}, followed by analysis in \cite{Alikakos1994} and \cite{Chen1996}. 
For the cases where the interfaces intersect an exterior boundary, which is the focus of our investigation, an additional condition at the contact line is required. We use the expression derived rigorously through a sharp-interface limit for the stationary Cahn--Hilliard equation by Modica~\cite{Modica1987a}.  The expression yields a Young-type condition for the contact angle in terms of the surface free energy contribution from the walls that closes the sharp-interface model. 

We further exploit the separation between the lateral and the vertical length scales of the sharp-interface profile to derive a new thin-film model for the free interface. This model greatly facilitates the systematic quantitative numerical study. It also enhances our understanding of the dynamics via the mathematical properties of the associated thin-film boundary value problem. This is used to discuss the bilayer breakup in Section 5. In Section 6, we give our conclusions and an outlook.

\section{Formulation of the phase field model}\label{sec:ch}

\paragraph*{Bulk equations}
For a mixture of two species, $A$ and $B$ that undergo phase separation below a critical temperature $T=T_c$ we introduce a phase-field model based on the Cahn--Hilliard equation. 
Besides the original work by Cahn and Hilliard \cite{Cahn1958} and by Cahn \cite{Cahn1961}, there is a vast original literature and reviews on such types of phase-field models including differences in the derivation and the scope of the modelling, e.g.\
\cite{Novick-Cohen2008,Nestler2011,Penrose1990,Gurtin1994}.
In our formulation, the phase-field parameter $\phi$ is a conserved
order parameter, obtained, for example, as a scaled volume or mole fraction,
where
$\mfnd=1$ represents the pure A-species and $\mfnd=-1$ the pure $B$ species, and $\mfnd=0$ a
symmetric, or 50:50 mixture of the two species. 
The nondimensional bulk equations in the domain $\Omega=\{(x,z):\,x\in\mathbb{R},\,0<z<d\}$ is given
by
\begin{subequations}\label{chndall}
\begin{align}\label{chnd}
\mfnd_t &= \nabla\cdot\left[(1-\mfnd^2)\nabla \mu\right],\\
\mu&=\frac1T \left[ f'(\mfnd)-{\eps^2}\Delta \phi\right], \label{chndb}\\
f(\mfnd)&=-\mfnd^2+ 
T \left[
(1-\mfnd)\ln(1-\mfnd)+(1+\mfnd)\ln(1+\mfnd)\right],
\label{ffnd}
\end{align}
\end{subequations}
where $\mu$ is the chemical potential and $T$ the temperature.
We will vary the temperature and therefore, the explicit dependence
on $T$ has been retained in \eqref{chndb}.
The parameter $\varepsilon$ is the ratio of the microscopic length scale of
the interaction between the two species---a quantity that can be expressed in terms
of the lattice parameter in the case of nearest neighbour interactions in a cubic lattice,
see, for example, \cite{Cahn1958}---and the macroscopic length scale use to nondimensionalise
the system. For the latter, we can assume, for example, the thickness of one of the layers
in the bilayer state that we will investigate has been scaled to one.

The boundary conditions are
\begin{subequations}\label{bcsnd}
\begin{alignat}{2}
\mu_z&=0, &\quad \eps\mfnd_z&=f_0'((1+\mfnd)/2)\quad \text{at } z=0,\\
\mu_z&=0, &\quad \eps\mfnd_z&=f_0'((1-\mfnd)/2)\quad \text{at } z=d,
\end{alignat}
\end{subequations}
where the left two conditions, $\mu_z=0$, correspond to no-flux through the substrate,
and the other two represent the interaction of the species with the substrate. 
The specific choice of $f_0$ is introduced further below. 
The chemical potential and the latter two boundary conditions arise as
the first variation of the free energy $F/T$ of the system, where the functional $F$ is 
given by
\begin{align}
  F[\mfnd] &=  \int_{0}^{d}\int_{-\infty}^{+\infty}
f(\mfnd(x,z)) + 
\frac{\varepsilon^2}{2} \left|\nabla \mfnd(x,z)\right|^2 \,\mathrm{d}x\mathrm{d}z
\notag
\\
 &\quad+2\varepsilon\int_{-\infty}^{+\infty}
  f_0((1+\mfnd(x,0))/2) \,\mathrm{d}x+2\varepsilon\int_{+\infty}^{-\infty} f_0((1-\mfnd(x,d))/2) \,\mathrm{d}x.
\notag
\end{align}
For $T$ below the critical temperature $T_c$, which here has been
scaled to one, the homogeneous contribution $f$ to the bulk has a
double-well structure and will drive the system to phase separate into
domains with compositions that correspond to the minima of $f$. The
choice of substrate-material interface energy density assumes antisymmetric substrates since
the integrand in the substrate integral at $z=0$ is transformed into
the integrand of the integral at $z=d$ if $\mfnd$ is replaced by
$-\mfnd$. Thus, the affinity of the upper substrate to species A is
the same as the affinity of the lower substrate to B. 
Using such interface energy distributions assumes that we only consider short-range 
surface interactions; other possibilities also include contributions
to the bulk free energy \cite{Puri1997b}.

We are mostly interested in phase-separating situations close to
criticality, with $T$ below and close to $1$, where the minima of
$f$, denoted by $\mfnd_{\pm}$, are close to zero.  In addition,
we intend to consider different choices for the temperature and in
Section~\ref{sec:bil}, we also prescribe time-dependent temperature
profiles.  Thus, we let $(1/T-1)=\chi_0 \chi(t)$ with a new parameter
$\chi_0\ll 1$, and a function $\chi(t)$, $0\leq \chi(t)\leq 1$.
It is then convenient to let
\begin{align}
  \phi = (3\chi_0)^{1/2} \hat{\phi}, \quad \mu = 3^{1/2} \chi_0^{3/2} \hat{\mu}, \quad t = \varepsilon^{-1}(\chi_0 + \chi_0^2)^{-1/2} \hat{t}, 
\end{align}
in \eqref{chndall}, \eqref{bcsnd},
so that, to leading order in $\chi_0\ll 1$, we obtain
\begin{subequations} \label{beapp2}
\begin{align}
 \hat\varepsilon\hat{\phi}_{\hat{t}} &= \Delta \hat{\mu}, \\
  \hat{\mu} &= \hat{f}'(\hat{\phi}) - \hat{\varepsilon}^2 \Delta \hat{\phi}, \\
  \hat{f}(\phi) &= \frac{1}{2} (\hat{\phi}^2 - \chi(t))^2,
\end{align}
where $\hat{\varepsilon} \equiv \varepsilon (1 +
1/\chi_0)^{1/2}$. We remark that in Section \ref{sec:sharp_interface} we will consider a sharp-interface limit
where $\hat{\varepsilon} \ll 1$ which effectively puts a lower bound on the value of $\chi_0$, namely $\varepsilon^2 \ll \chi_0$.
Furthermore, $\hat{f}$ has been altered
by a $\phi$-independent function of time
which is immaterial here.
The corresponding rescaled boundary conditions are
\begin{alignat}{2} \label{bcsend2}
  \hat{\mu}_z &= 0, &\quad
  \hat{\varepsilon} \hat{\phi}_z &= \hat{\beta}(1 - \hat{\phi}^2)
\quad \text{at } \hat{z}=0,\,d,
\end{alignat}
\end{subequations}
where we have made a specific choice for $\hat f_0$, 
\begin{equation}
\hat f_0(\hat \mfnd)=\hat\beta (\hat\mfnd-\hat\mfnd^3/3), \label{eqn:f0}
\end{equation}
so that the surface energies at $z=0$ and $z=d$ are now $\hat f_0(\hat
\mfnd)$ and $\hat f_0(-\hat \mfnd)$, respectively.  Other typical
choices for $\hat{f}_0$ involve quadratic polynomials \cite{Puri1994}.
For the determination of the effective surface energies and the contact
angle when $\chi=1$ (see Section~\ref{sec:siltfa}) it is more convenient \cite{xu_analysis_2011} to use the
expression \eqref{eqn:f0} above for which the derivative
of $\hat{f}_0$ vanishes at the minima of the bulk free energy $\hat f$.

In the following, we drop the hats from all variables and parameters.

\section{Formation of bilayers}\label{sec:bil}

Using the model developed in the previous section, we can
now investigate the conditions in which the two constituent
components of the mixture separate and form a horizontal bilayer. This
investigation will be guided by linear stability results and numerical simulations.  
We assume that the initial condition is a small random perturbation with mean value zero
to the homogeneous 50:50 state (the latter of which corresponds to $\phi \equiv 0$).
We typically set $\varepsilon = 0.127$ and for the numerical simulations, the 
domain is truncated at $x=0$ and $x=\lcut$
with $\lcut\gg d$, and we impose periodic boundary conditions.

For definiteness and to facilitate the discussion we interpret the function $\chi(t)$ as the 
temperature of the system in the remaining part of our study. The arguments, however, are
 general and can equally be made for other realisations, such as the concentration of the
 species. 

\paragraph{Stability analysis}
We investigate the linear stability of one-dimensional stationary solutions
when the temperature is held at a constant value. Thus, we set $\chi(t) \equiv \chi$
and write the order parameter and the chemical potential as
\begin{subequations}
\label{eqn:lin_ansatz}
\begin{align}
  \phi(x,z,t) &= \bar{\phi}(z) + \alpha \tilde{\phi}(z)e^{\lambda t + ikx}, \\
  \mu(x,z,t) &= \bar{\mu}(z) + \alpha \tilde{\mu}(z) e^{\lambda t + ikx},
\end{align}
\end{subequations}
where bars are used to represent the stationary solution and tildes denote
perturbations to it. The parameters $\alpha \ll 1$, $\lambda$, and $k$
denote the initial amplitude of the perturbation and its
growth rate and wavenumber, respectively. The solution 
\eqref{eqn:lin_ansatz} is inserted into the governing equations and their
boundary conditions \eqref{beapp2}, and the system is expanded in powers of $\alpha$. 

The $O(1)$ contribution to this system describes the steady, one-dimensional problem.
From this we find that the chemical potential satisfies $\bar{\mu}_{zz} = 0$ with 
$\bar{\mu}_z = 0$ on the boundaries. Therefore, the chemical potential is constant
to leading order and we write $\bar{\mu}(z) \equiv \bar{\mu}$. 
The problem for the order parameter can be written as
\begin{subequations}
\begin{align}
  \bar{\mu} = f'(\bar{\phi}) - \varepsilon^2\bar{\phi}_{zz},
\end{align}
with boundary conditions
\begin{align}
  \varepsilon \bar{\phi}_z = \beta\left(1 - \bar{\phi}^2\right), \quad z = 0, d.
\end{align}
The chemical potential is treated as a Lagrange multiplier that ensures the steady solution
corresponds to a 50:50 mixture; thus, we supplement the boundary value problem with the integral
condition given by
\begin{align}
  \int_{0}^{d} \bar{\phi}(z)\,\mathrm{d} z = 0.
\end{align}
\end{subequations}

The stability of the stationary solution is determined from the $O(\alpha)$ problem
which can be written as
\begin{subequations}
\begin{align}
  \varepsilon\lambda \tilde{\phi} &= -k^2 \tilde{\mu} + \tilde{\mu}_{zz}, \\
  \tilde{\mu} &= f''(\bar{\phi})\tilde{\phi} - \varepsilon^2(-k^2\tilde{\phi} + \tilde{\phi}_{zz}),
\end{align}
with boundary conditions
\begin{align}
  \tilde{\mu}_{z} = 0, \quad \varepsilon \tilde{\phi}_{z} &= -2 \beta\,\bar{\phi}\,\tilde{\phi}, \quad z = 0, d.
\end{align}
\end{subequations}

Given that the steady solution $\bar{\phi}$ is generally a function of space, the linear
stability problem is non-autonomous and can only be solved in exceptional circumstances. 
Such is the case when
the substrate-material interface energy is neglected, i.e., when $\beta = 0$,
or when it is very small, $\beta \ll 1$. 
In both instances the steady solution
is given by $\bar{\phi} \equiv 0$ (or to leading-order in the latter case) 
and the perturbations take the form
\begin{align}
  \tilde{\phi}(z) = \tilde{\phi}_n \cos(n \pi z / d),
  \quad
  \tilde{\mu}(z) = \tilde{\mu}_n \cos(n \pi z / d),
\end{align}
where $n \in \mathbb{Z}$. The linearised problem can be solved and the growth rates
are found to be
\begin{align}
  \lambda(\xi) = -\varepsilon^{-1}\xi^2 \left(\varepsilon^2\xi^2 - 2\chi \right),
\end{align}
where $\xi^2 = k^2 + (n \pi / d)^2$ represents an average wavenumber that
is composed of a continuous horizontal wavenumber $k$, and a discrete
vertical wavenumber $n$. Perturbations with average wavenumbers that
satisfy
\begin{align}
  0 < \xi^2 < \xi_c^2 = \frac{2 \chi}{\varepsilon^2},
\end{align}
have positive growth rates and hence the one-dimensional stationary solution
is linearly unstable. The fastest-growing perturbations have wavenumbers
that satisfy $\xi_m = \xi_c / \sqrt{2}$ and these lead to the formation
of distinct domains that are rich in one particular phase. The initial size
of these domains is approximately equal to half of the wavelength
of the most unstable modes. Over time these domains will coarsen
until the total interfacial area between them reaches a minimum. Due to
the assumption of the film being longer than it is wide, it becomes
energetically favourable for the system to form domains that
resemble a series of vertical columns instead of a horizontal bilayer. 

\begin{figure}[p]
\hspace{2pt}
\subfigure[]{\includegraphics[clip=true, width=0.330\textwidth]{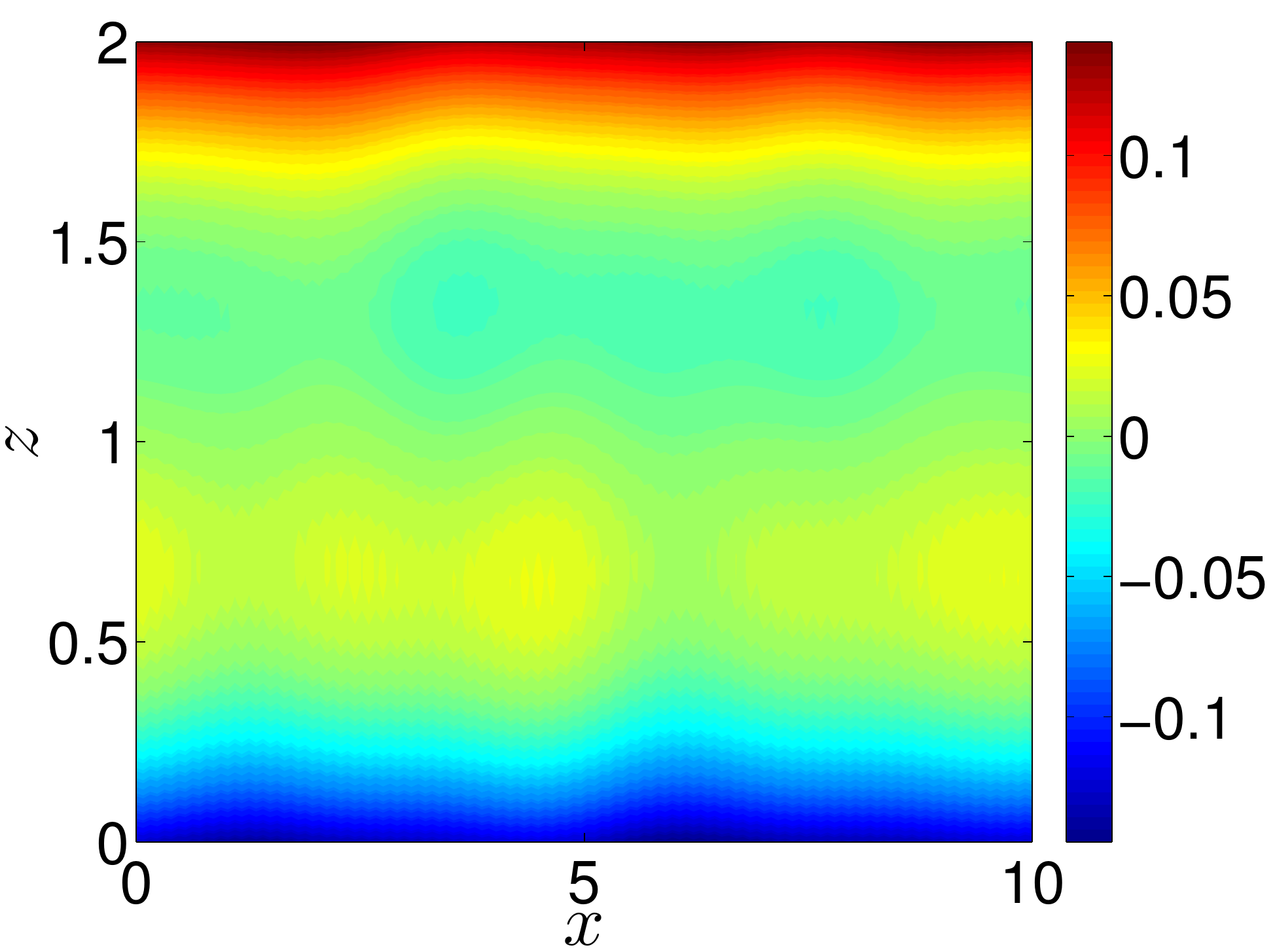}}
\subfigure[]{\includegraphics[clip=true, width=0.320\textwidth]{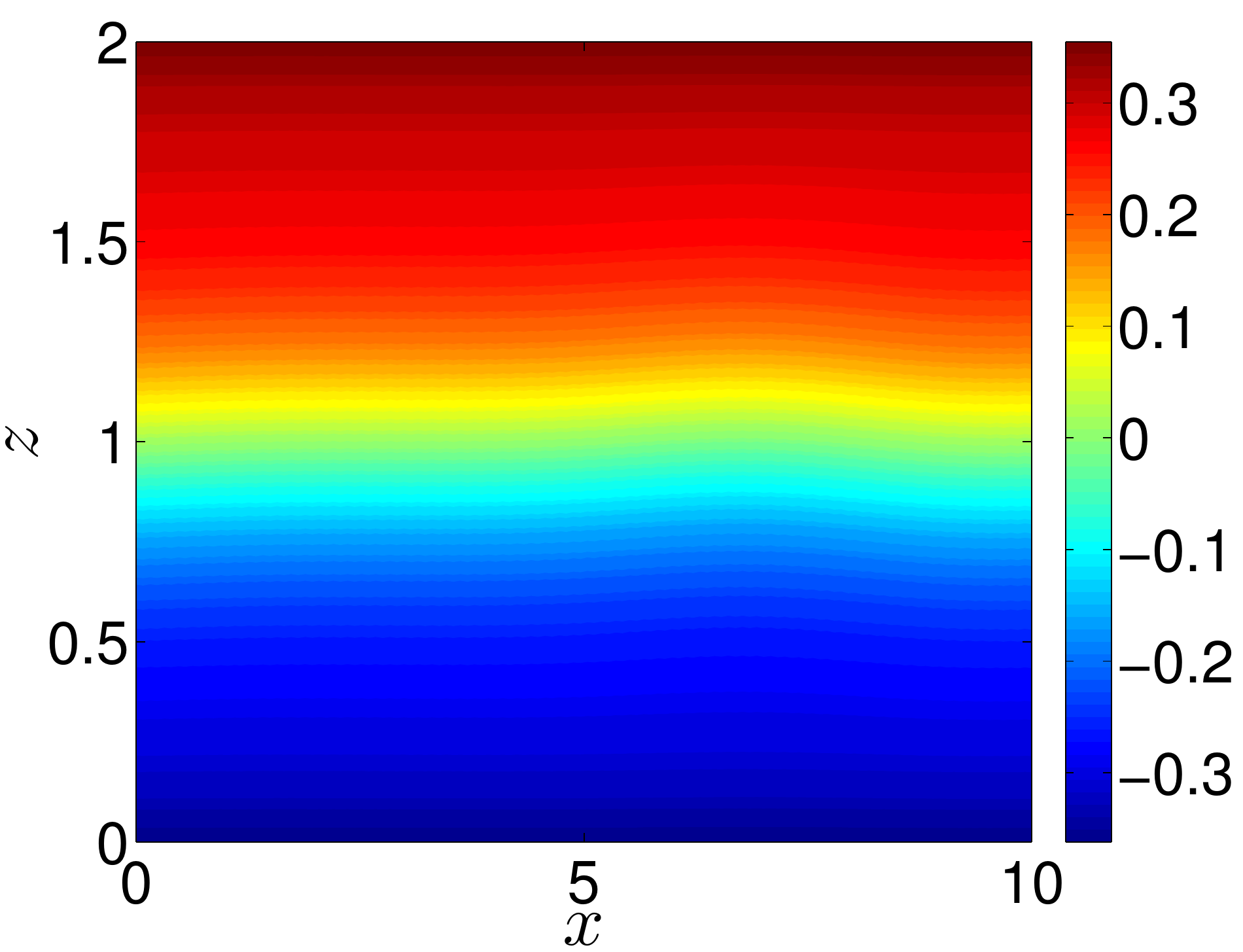}}
\subfigure[]{\includegraphics[width=0.320\textwidth, clip, trim=0 0 0 0]{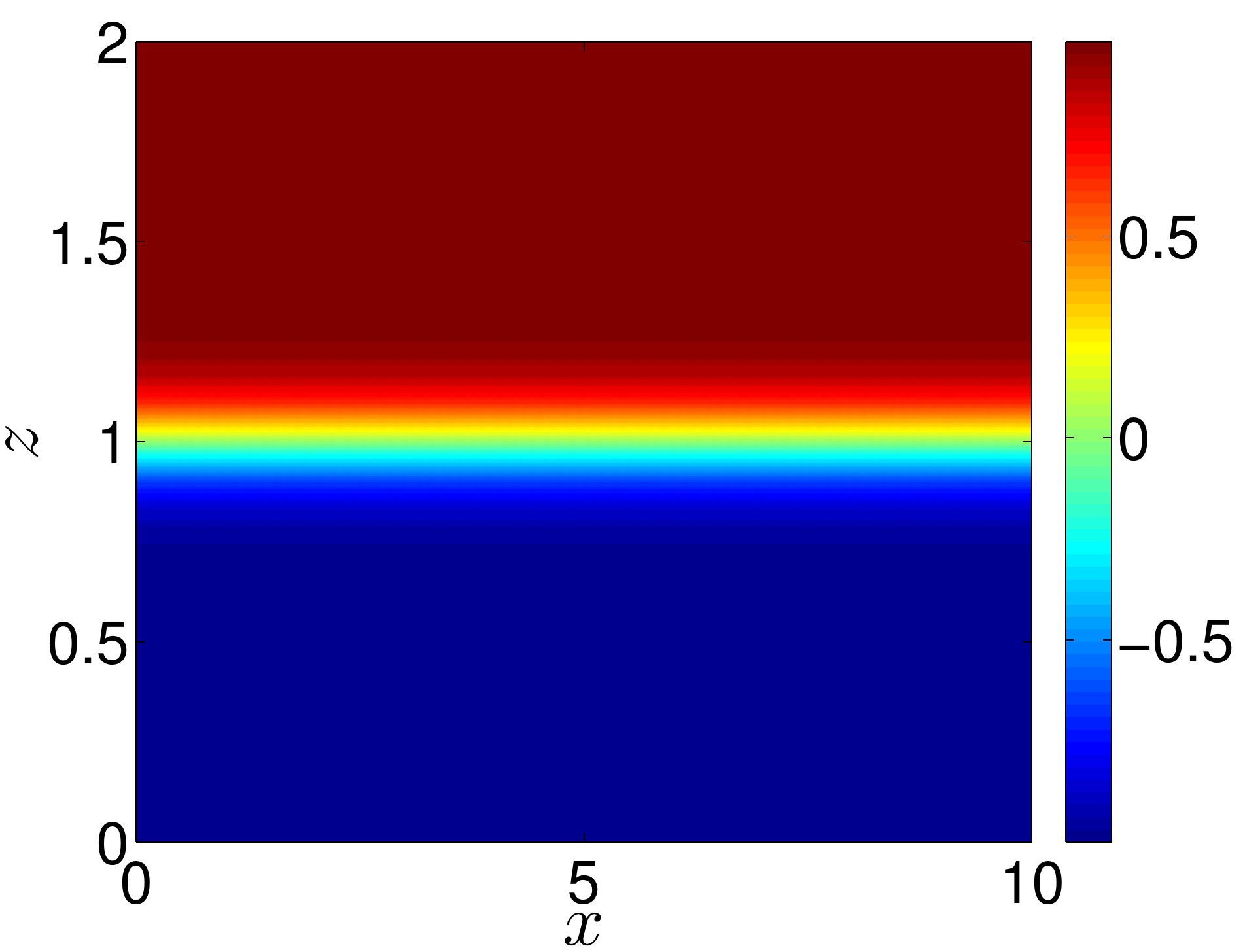}}
\\
\subfigure[]{\includegraphics[clip=true, width=0.32\textwidth]{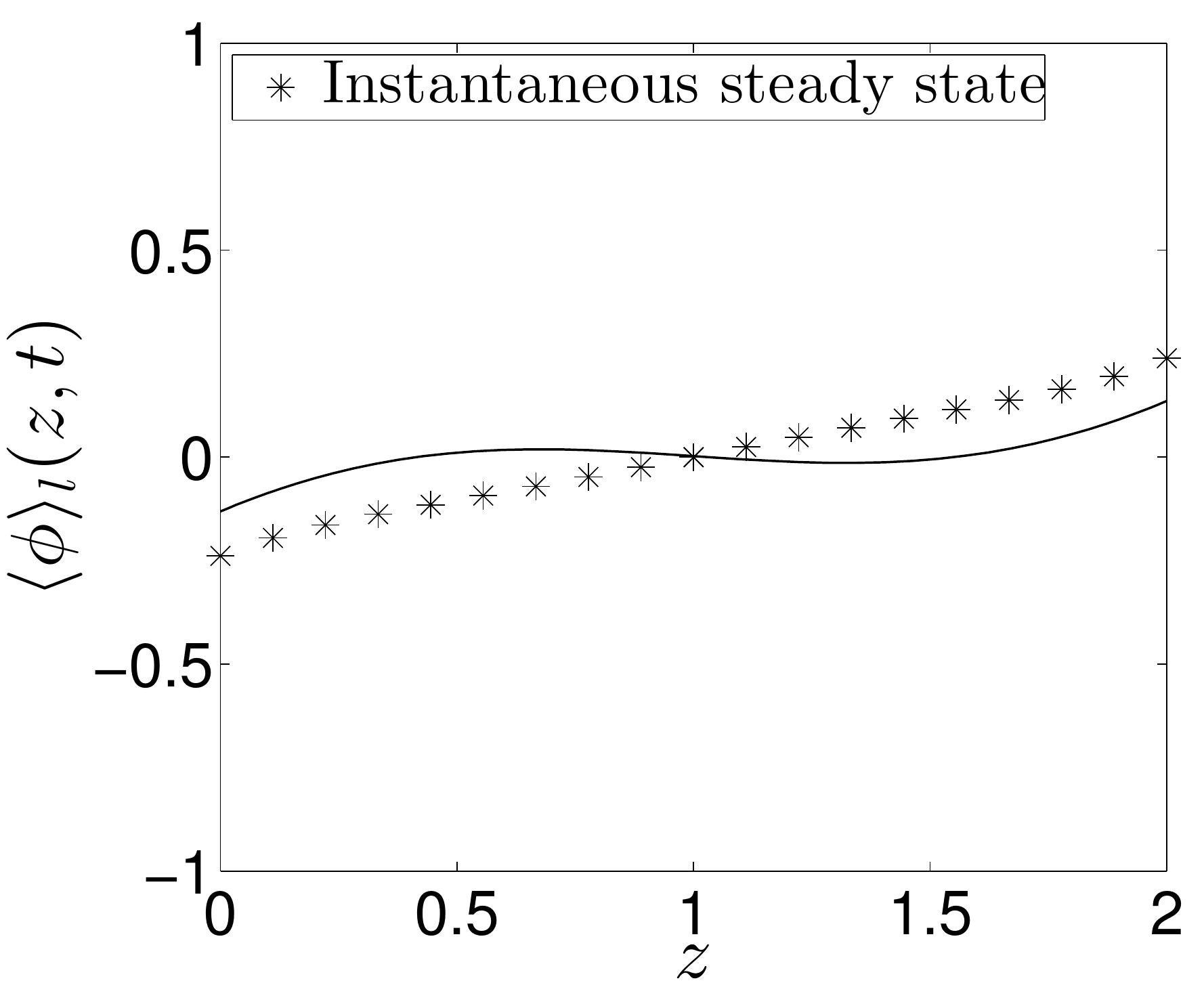}}
\subfigure[]{\includegraphics[clip=true, width=0.32\textwidth]{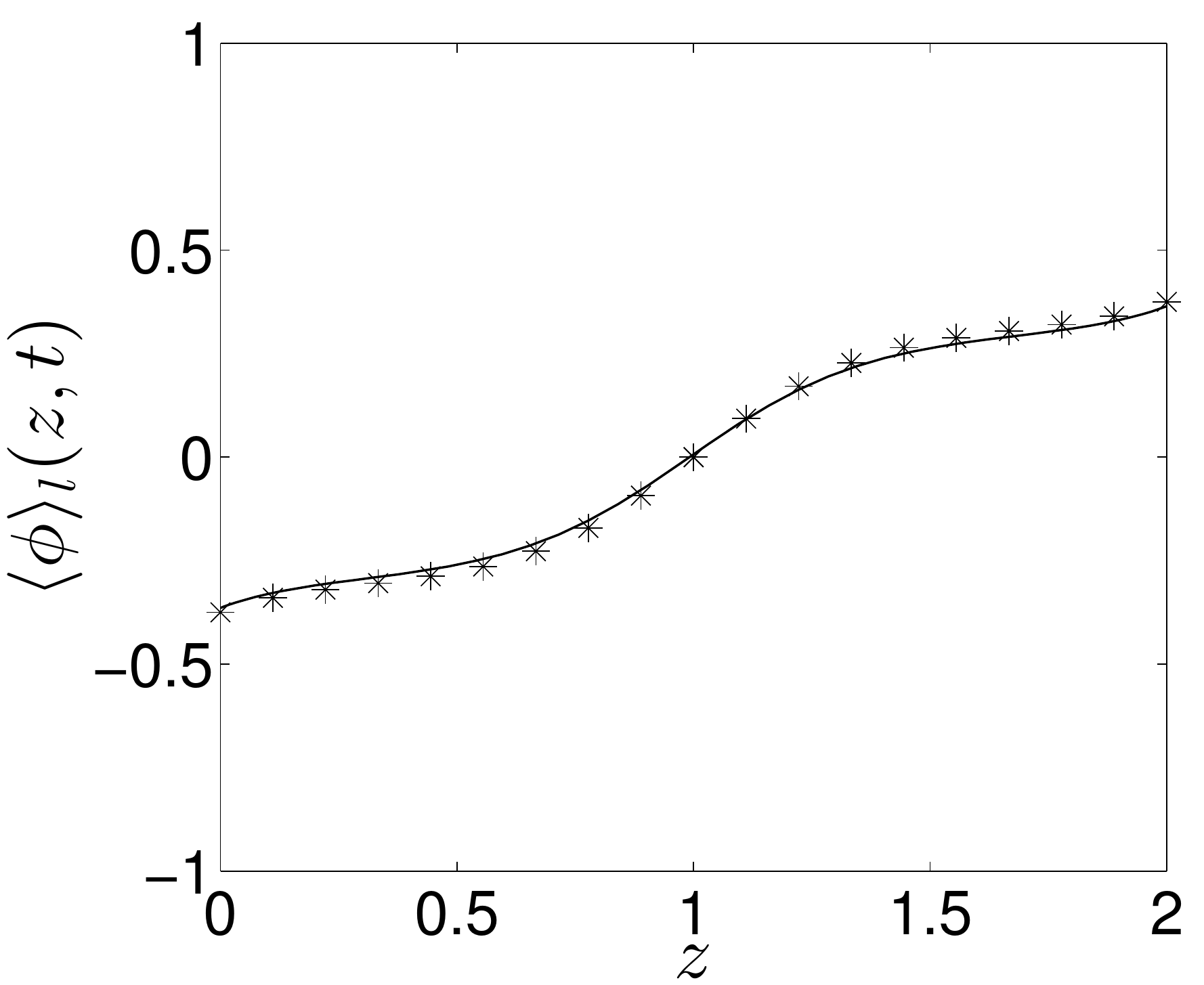}}
\subfigure[]{\includegraphics[clip=true, width=0.32\textwidth]{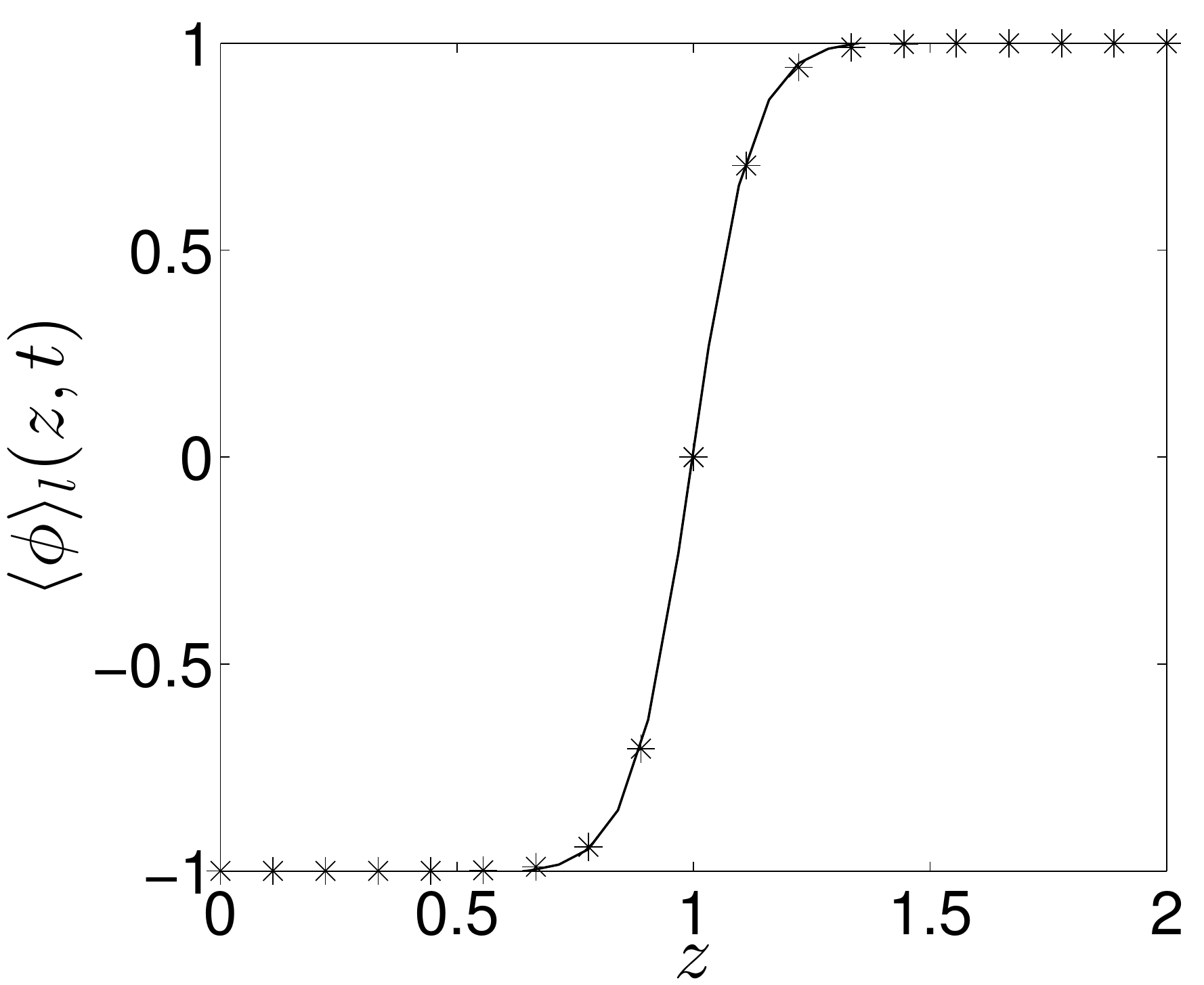}}
\\
\subfigure[]{\includegraphics[clip=true, width=0.32\textwidth]{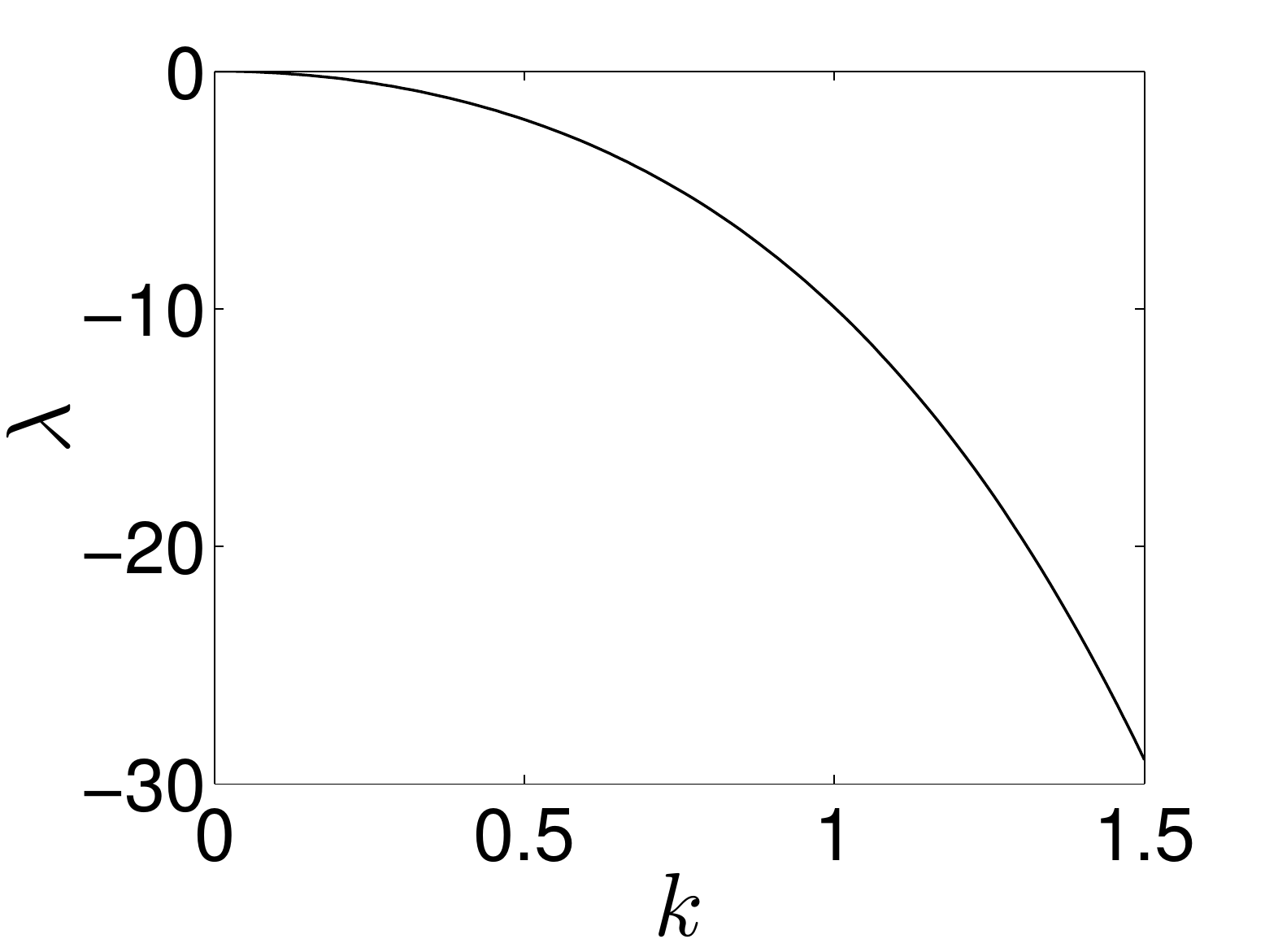}}
\subfigure[]{\includegraphics[clip=true, width=0.32\textwidth]{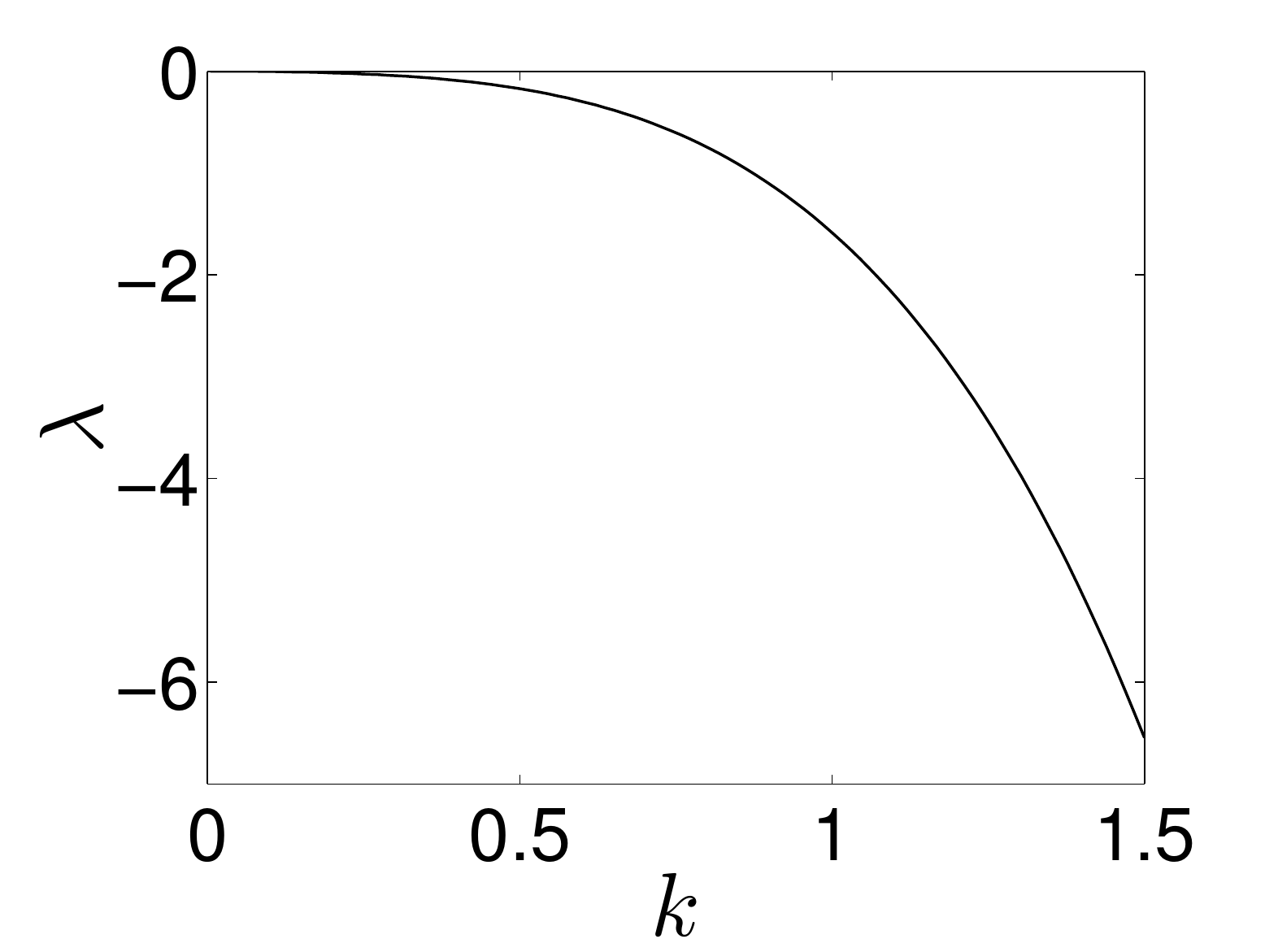}}
\subfigure[]{\includegraphics[clip=true, width=0.32\textwidth]{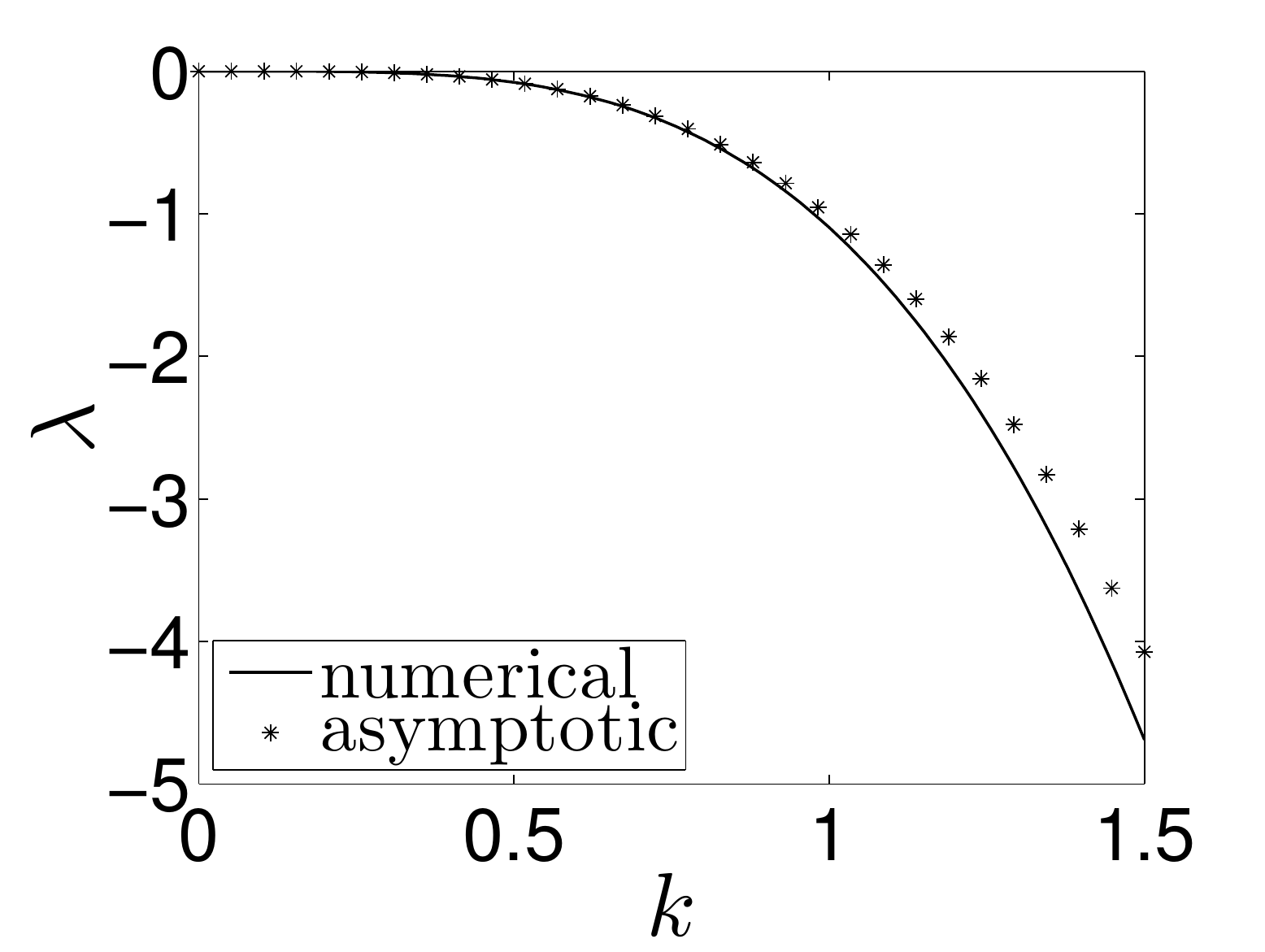}}
\caption{Bilayer formation when the temperature of the system is slowly
decreased according to the function $\chi(t) = 1 - \exp(-t/\tau_c)$. The initial condition was 
a randomly-perturbed homogeneous 50:50 mixture. 
Top (a)--(c): Evolution of the order parameter. 
The solution is shown at times $t = $ 0.17, 1.7, and 169, corresponding to
$\chi(t) = $ 0.01, 0.1, and 1, respectively.
Middle (d)--(f): Comparison of the laterally-averaged
order parameter $\langle \phi \rangle_l$ (solid line) with
the instantaneous
stable steady state solution (stars) at the same times as in
panels (a)--(c).
Bottom (g)--(i): The growth rate $\lambda$ of perturbations to the
steady state solutions shown above as functions of the perturbation
wavenumber $k$. See text for the definition of
$\langle \phi \rangle_l$ and the specific choice of $\tau_c$ that
was used.
}
\label{fig:cooling}
\end{figure}

When $\beta \neq 0$, the energetic interactions between the substrates and the constituent
materials induce a layered morphology in the steady-state solution
of the one-dimensional problem. Examples of these solutions have been computed numerically
and are shown as stars
in Fig.~\ref{fig:cooling} (d)--(f) when $\beta = 0.063$ and for temperatures
given by $\chi = 0.01$, $0.1$, and $1$, respectively. For 
temperatures below, but close to criticality (panel (d)), the steady state solution
resembles a small, approximately linear perturbation to the homogeneous 50:50 state.
For cooler temperatures, the solutions correspond to bilayer configurations
(panels (e)--(f)). By solving the corresponding linear stability problem numerically
as well, we find that these steady states are linearly stable; that is, the growth 
rates are negative for each value of the perturbation wavenumber $k$. The functional
forms of $\lambda$, i.e., $\lambda = \lambda(k)$, are shown in Fig.~\ref{fig:cooling} (g)--(i).

In the case when  $\varepsilon \ll 1$ and $\chi = 1$, the stability
problem can be solved by matched asymptotic expansions. The leading-order composite solution
to the stationary problem corresponds to a bilayer and is given by
\begin{subequations}
  \begin{align}
    \label{eqn:tanh}
  \bar{\phi}(z) = \tanh\left(\frac{z - 1/2}{\varepsilon}\right).
\end{align}
The perturbation to the order parameter and its growth rate are, to leading order, given by
\begin{align}
  \tilde{\phi}(z) &= A\,\mathrm{sech}^2\left(\frac{z - 1/2}{\varepsilon}\right), \\
  \lambda(k) &= -\frac{2}{3}\,k^3\tanh(k/2),
\end{align}
\end{subequations}
respectively, where $A$ is a multiplicative constant. A comparison of the asymptotic
and numerical growth rates is shown in Fig.~\ref{fig:cooling} (i). There is good
agreement between the two, particularly when the wavenumber $k$ is small.

The linear stability of the bilayer configuration over a wide range of 
temperatures suggests that a robust method for driving the system
into such a state is to slowly cool the system from a near-critical
temperature. By starting from a temperature close to the critical value,
the influence of the substrates will induce a layered morphology and 
push the mixture towards its stable steady state profile. Decreasing
the temperature at a sufficiently slow rate will then allow the mixture to
evolve in a quasi-stationary manner that follows the stable steady state
profile, thus yielding a bilayer configuration for cooler temperatures.

\begin{figure}[p]
\centering
\subfigure[]{\includegraphics[width=0.4\textwidth]{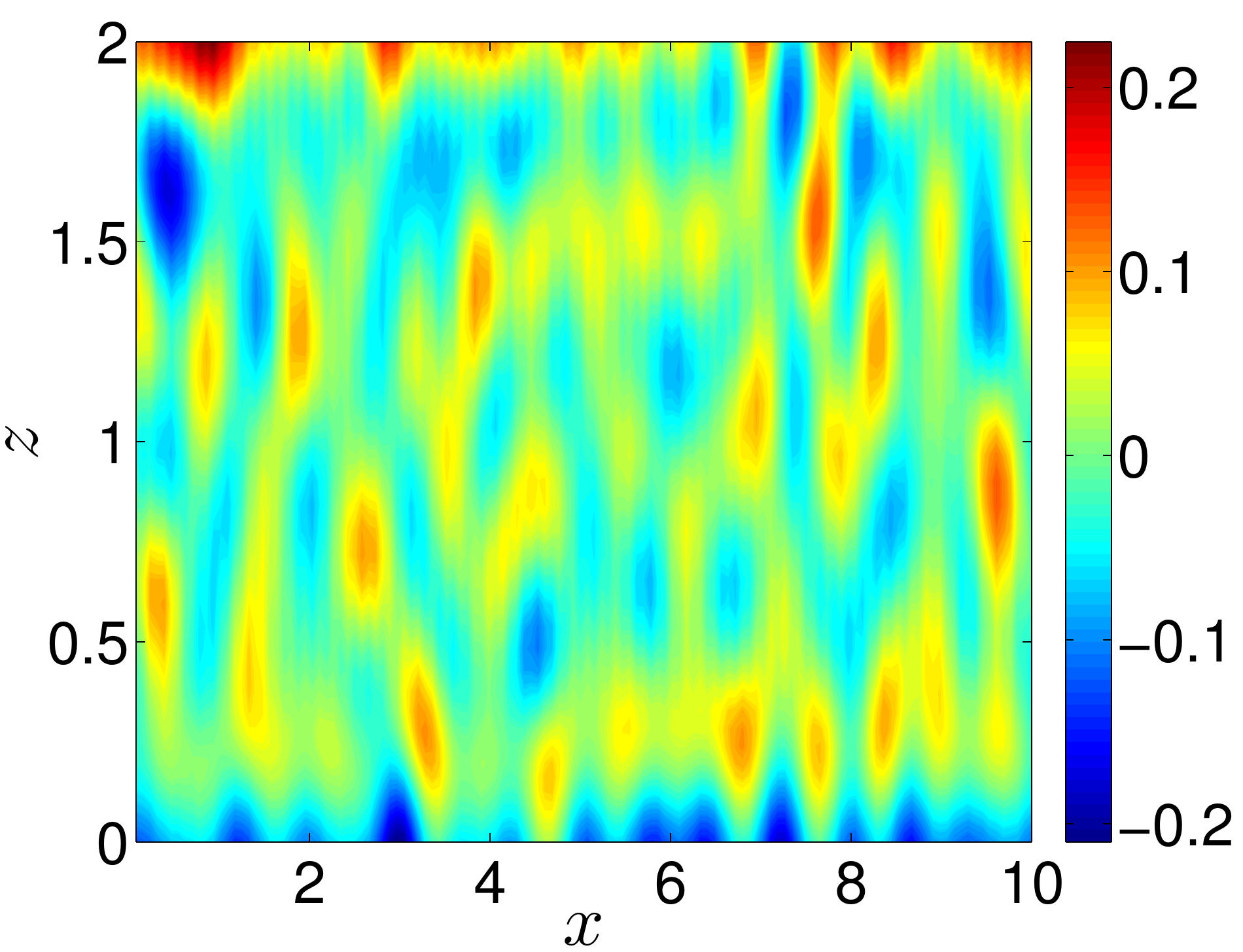}}
\subfigure[]{\includegraphics[width=0.4\textwidth]{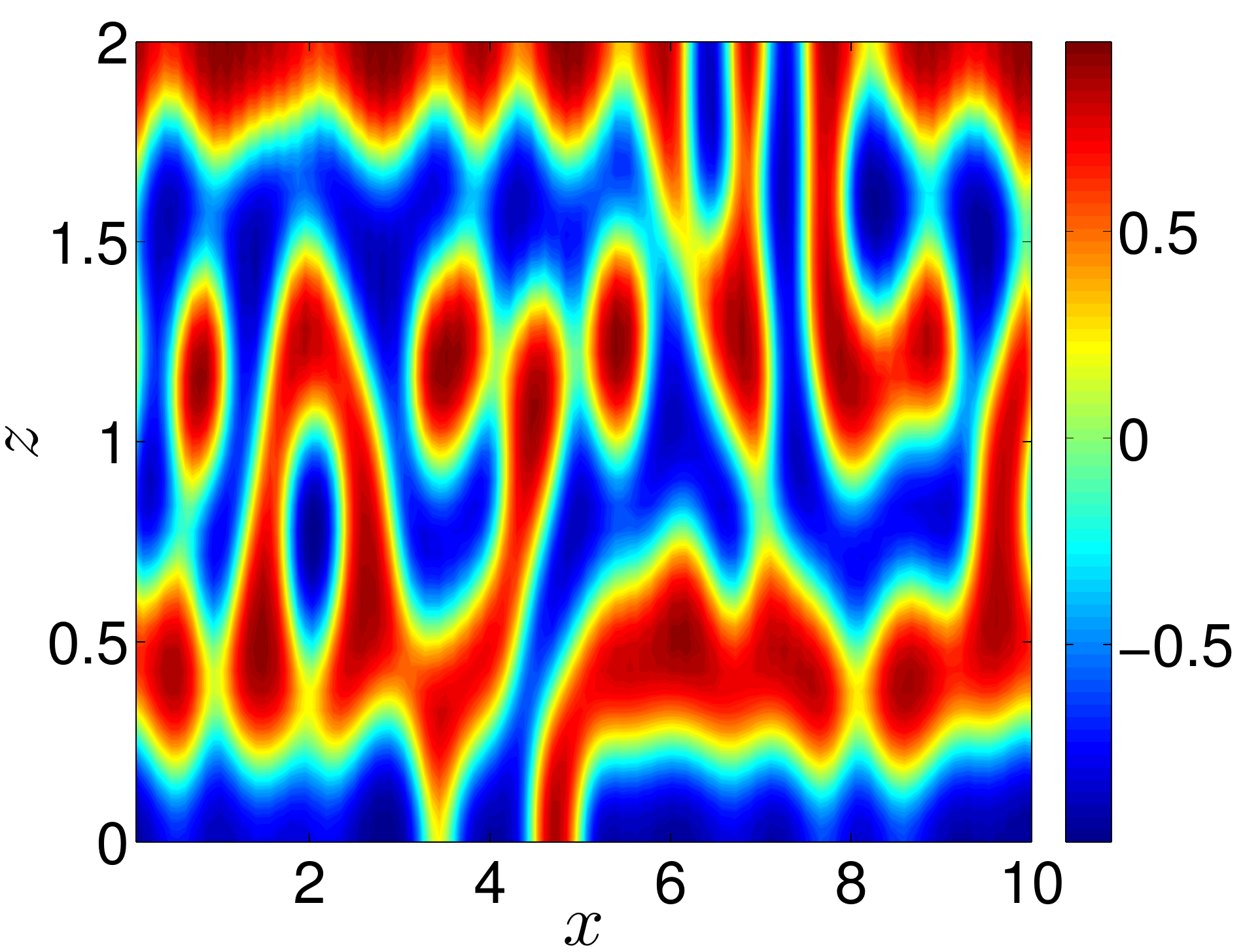}} \\
\subfigure[]{\includegraphics[width=0.4\textwidth]{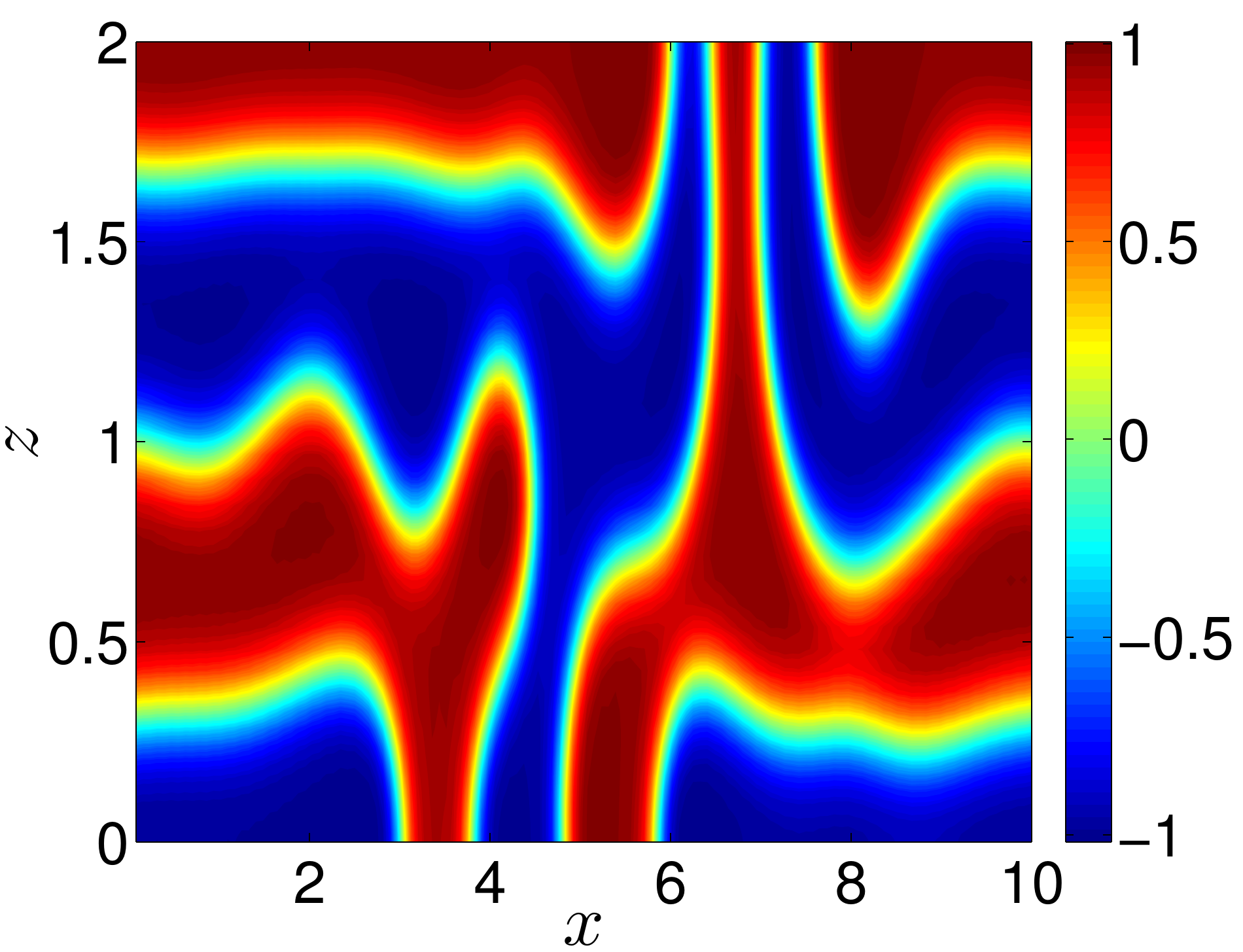}}
\subfigure[]{\includegraphics[width=0.4\textwidth]{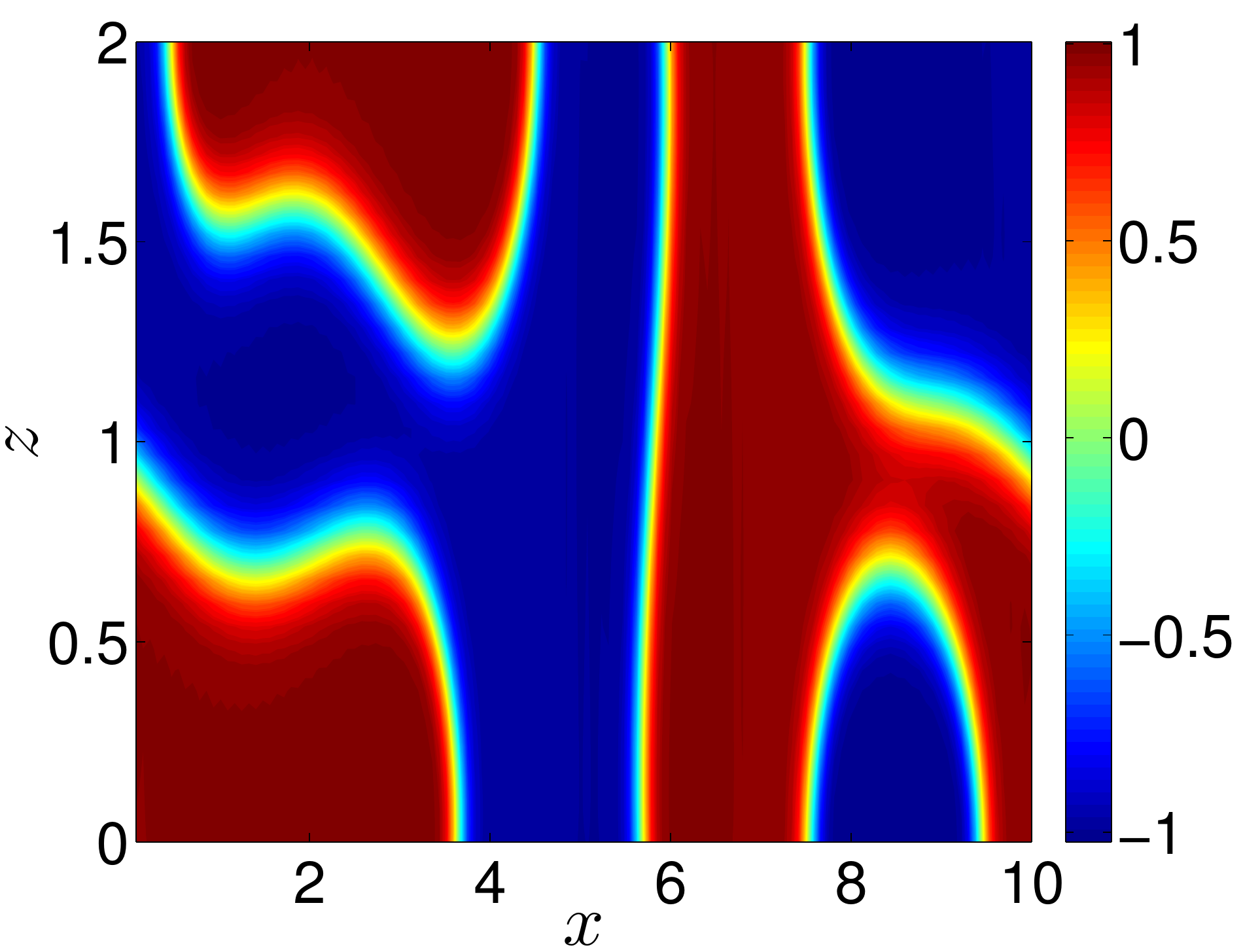}} \\
\subfigure[]{\includegraphics[width=0.4\textwidth]{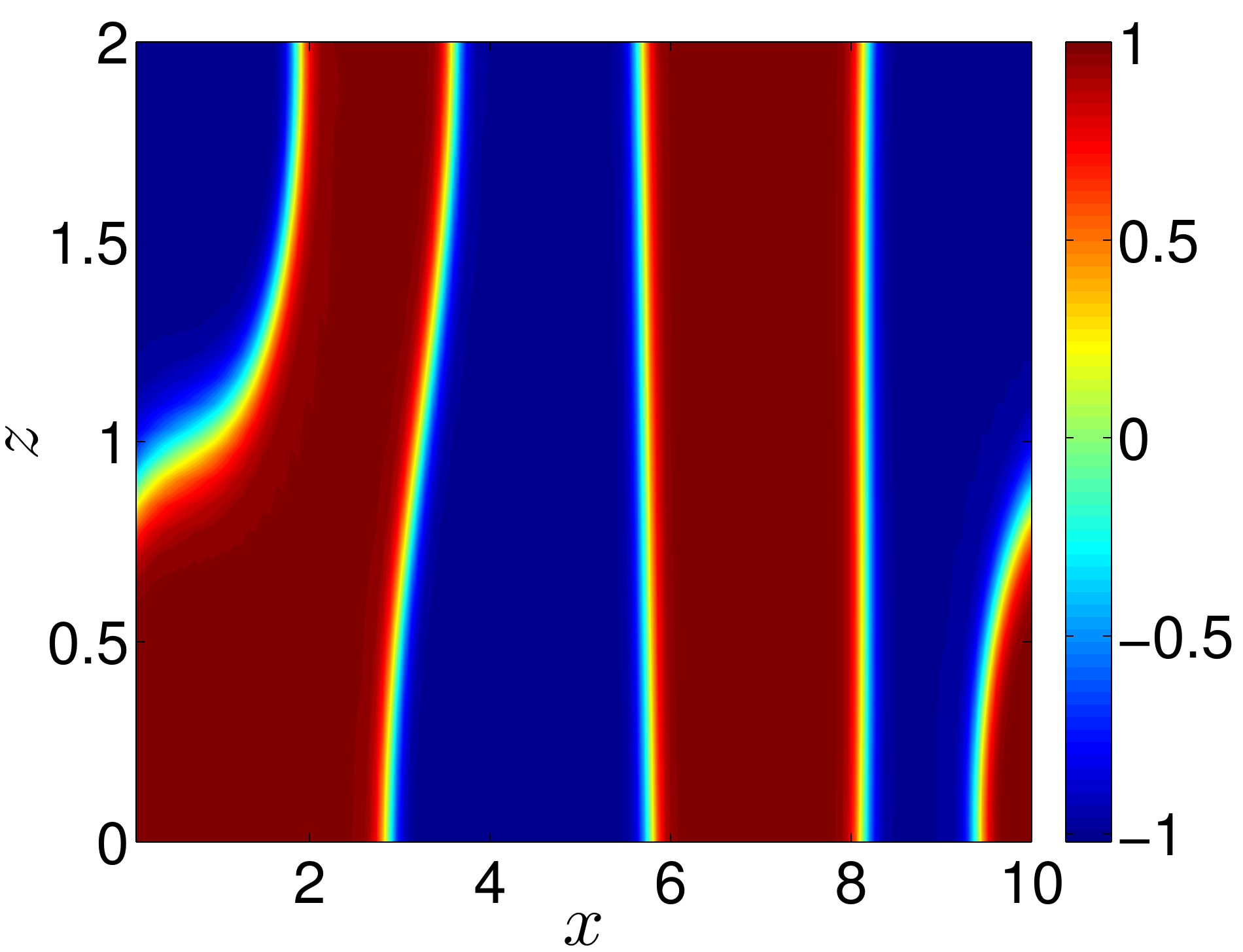}}
\subfigure[]{\includegraphics[width=0.4\textwidth]{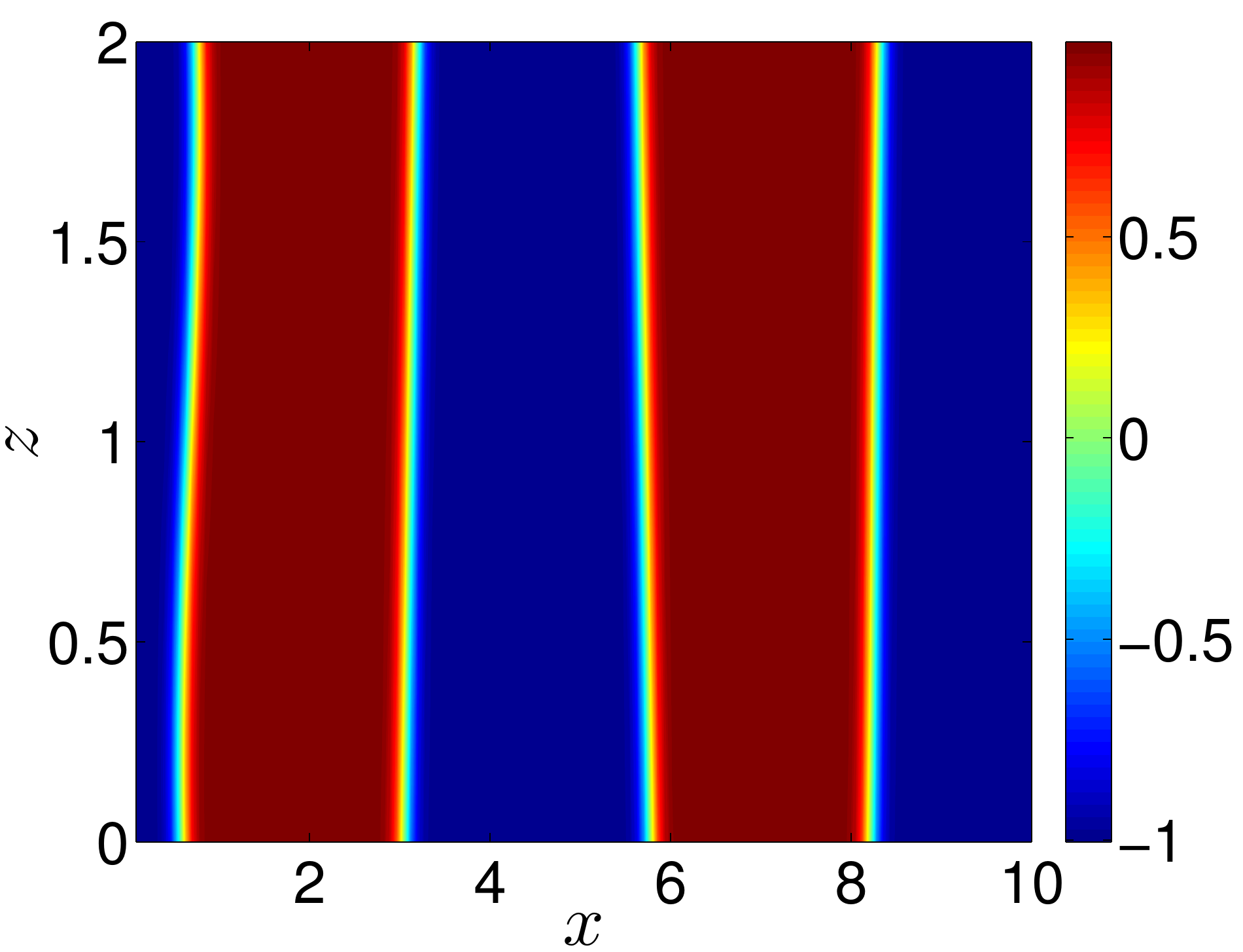}}
\caption{Spinodal decomposition and coarsening in a system held at a constant
  temperature below the critical value. The temperature was fixed 
  at $\chi \equiv 1$ and the initial condition was a randomly-perturbed
  homogeneous 50:50 mixture. The solution is shown
at times 
$t=1.7\times 10^{-3}$, $1.7\times 10^{-2}$, $0.084$, $0.84$, $1.7$, and $2.8$.}
\label{fig:constant_temp}

\end{figure}

\afterpage{\clearpage}

\paragraph{Numerical solution of the phase-field model}

The numerical simulations are
based on an implicit-explicit spectral method. The time derivative
is discretised using the standard first-order finite difference
approximation and any linear terms are handled implicitly, whereas
nonlinear terms are treated explicitly. The solutions are assumed to be
periodic in the horizontal direction and hence derivatives with respect
to $x$ are computed using Fourier spectral methods. Chebyshev spectral
methods are used to compute derivatives in the vertical direction.

Using this numerical scheme, we first explore the dynamics that occur when 
the temperature of the system is fixed at a constant value that is below 
criticality. The interface energy between the substrates and the components is also taken into consideration.
Thus, we set $\chi \equiv 1$ and $\beta = 0.063$.  The computational
domain is $L_\infty = 10$, and the initial condition is a random perturbation
of amplitude 0.2 to the homogeneous 50:50 state. The simulations results,
which are shown in Fig.~\ref{fig:constant_temp}, indicate that the 
initial fluctuations in the solution are rapidly amplified by
spinodal decomposition, producing small domains that are nearly pure 
in the two constituent species. The width of these domains is approximately
0.4, which is in good agreement with the size that is predicted from
the linear stability analysis; in this case the wavenumber of the 
fastest growing mode is $\xi_m \simeq 7.9$. 
For $t > 1.7\times 10^{-2}$ these domains coarsen to form large-scale structures that eventually
settle into columns.

We now investigate how the system evolves when it is slowly cooled from the critical
temperature. In particular, the temperature is decreased according to
$\chi(t) = 1 - \exp(t/\tau_c)$ with $\tau_c = 16.9$. The values of
the other parameters are the same as in Fig.~\ref{fig:constant_temp};
we take $\beta = 0.063$, $L_\infty = 10$, and the initial noise has
an amplitude of 0.2. The results of the simulation are presented in 
Fig.~\ref{fig:cooling}. Panels (a)--(c)
of the figure display the evolution of the order 
parameter, whereas panels (d)--(f) compare, at various times, the laterally-averaged 
order parameter, defined as
\begin{align}
  \langle \phi \rangle_l (z,t) = \frac{1}{L_\infty} \int_{0}^{L_\infty} \phi(x,z,t)\,\mathrm{d} x,
\end{align}
to the instantaneous, stable steady-state solution of the one-dimensional problem.
The figure clearly shows that the cooling procedure is able to produce the bilayer
morphology for parameter values that lead to a columnar topology when the temperature
was held at a fixed value. The comparison of the laterally-averaged order parameter
and the stable steady state shows that the solution quickly adopts the steady-state
profile and evolves in a quasi-stationary manner to the bilayer state. 

The system will also tend to a bilayer configuration if the substrate-material
interface energy is high. However, in this case it is possible
for the bilayer state to minimise globally the energy of the system
and hence it would be expected to be stable. That is, no topological
transition could be initiated by nucleating a hole in the bilayer. This
situation relates to the substrates being perfectly wetting, so that any
hole in the bilayer will close up. 

\section{Asymptotic approximations}\label{sec:siltfa}

\subsection{Sharp-interface limit}
\label{sec:sharp_interface}
For the topological transition, we will consider the system
at a fixed temperature corresponding to $\chi = 1$. 
At this temperature the width of the diffuse interface between A-rich and B-rich
domains is $O(\varepsilon)$. Thus, for $\varepsilon \ll 1$, the thickness of the
transition layer is small and the phase-field model \eqref{beapp2} can be reduced to
a sharp-interface model as described, for example, by Pego \cite{pego_front_1989}. 
The order parameter and the chemical potential are written as an asymptotic series
of the form
\begin{subequations}
\begin{align}
  \phi &= \phi_0 + \varepsilon \phi_1 + O(\varepsilon^2), \\
  \mu &= \varepsilon \mu_1 + O(\varepsilon^2). \label{eqn:matt_expansions}
\end{align}
\end{subequations}
The leading-order solution for the order parameter is $\phi_0 = \pm 1$. The $O(\varepsilon)$
problem is given by
\begin{subequations}\label{mshat}
\begin{align}
  \Delta \mu_1 &= 0,\label{mshata}
\end{align}
in the regions $0 < z < h(x,t)$ and $h(x,t) < z < d$. Along the sharp interface $z = h(x,t)$ 
we have the boundary conditions
\begin{align}
  2\mu_1 &= \frac{\sigma h_{xx}}{(1 + h_x^2)^{3/2}}, \label{mshatb}\\
  h_{t} &= \frac{1}{2}\left([\mu_{1,x}]^{+}_{-} h_x - [\mu_{1,z}]^{+}_{-}\right). \label{mshatc}
\end{align}
 The parameter $\sigma$ is defined through the expression
\begin{align}
\sigma = \int_{-\infty}^{\infty} (\Phi_{0,\eta\eta})^2\, \mathrm{d} \eta = \frac{4}{3},
\end{align}
where $\Phi_0$ is the solution to the leading-order inner problem given by $2(\Phi_{0}^3 - \Phi_0) = \Phi_{0,\eta\eta}$ subject to $\Phi_0 \to \pm 1$ as $\eta \to \pm \infty$ and $\Phi_0 = 0$ at $\eta = 0$. The solution is given by $\Phi_0 = \tanh\eta$ which can be directly used to show $\sigma = 4/3$. 

Along the substrates $z = 0$ and $z = d$ we have the conditions 
\begin{equation}
\mu_{1,z} = 0. \label{mshatd}
\end{equation}
\end{subequations}


The corresponding leading-order composite solution for a known sharp interface $h(x)$ is useful to know, in particular, when constructing initial conditions for the phase-field model. This solution can be written as
\begin{align}
  \phi(x,z) = \tanh\left(\frac{z - h(x)}{\varepsilon\sqrt{1 + h_x^2}}\right),
  \label{eqn:si_comp_soln}
\end{align}
which is a generalisation of the expression presented in \eqref{eqn:tanh}. 

Where the interface touches a substrate, a condition for the contact angle is needed.
In the limit considered here, the system is nearly in equilibrium, so we use
the equilibrium contact angle which can be expressed in terms of the surface
and interface energies via a Young-Laplace formula,%
\begin{subequations}\label{scacond}
\begin{align}
  \cos\theta = \frac{\gamma(1) - \gamma(-1)}{\bar{\sigma}},
\end{align}
with
\begin{align}
  \bar{\sigma} = \int_{-1}^{1} (2 f(r))^{1/2}\, \mathrm{d}r = \sigma = \frac{4}{3},
  \end{align}
and
where the appropriate surface energies
need to include contributions from boundary layers near the substrates \cite{Cahn1977},
\begin{align}
  \gamma(\rho) = \inf_{\omega} \left\{ f_0(\omega) + \left| \int_{\omega}^{\rho} (2 f(r))^{1/2}\,\mathrm{d}r \right|\right\}.
\end{align}
\end{subequations}
A rigorous proof of \eqref{scacond} is given in  \cite{Modica1987a}.
If $\beta \leq 1$ one finds that $\gamma(\pm 1) = f_0(\pm 1) = \pm 2{\beta}/3$.
The expression for the contact angle reduces to
\begin{align}
  \label{eqn:cos_theta}
  \cos \theta = \beta.
\end{align}

Since we have assumed that the interface is given as the graph
of a function $z=h( x, t)$, we need to restrict $0\leq \theta\leq\pi/2$. 
The conditions at the contact line $ x= s( t)$ are
\begin{align}
 h&\to 0,\quad
 h_{ x} \to \tan\theta,\quad
 q \to 0,
\quad \text{as } x \to s(t). \label{sc}
\end{align}
The first condition is obvious and the last is a no-flux condition
that ensures that no mass is
lost through the contact line.  To the far right, the film flattens
to a constant film thickness and there is no flux,
\begin{align}
\label{ffc}
 h\rightarrow 1, \quad
 q\rightarrow 0, \quad \text{as }  x\rightarrow\infty.
\end{align}
Notice that the thickness of a uniform layer
is fixed to one by our choice of scalings (see Section \ref{sec:ch}).

We conclude this section with two useful mass conservation properties
for the sharp-interface model.  First, if we integrate \eqref{mshata}
twice with respect to $z$ and use \eqref{mshatb}--\eqref{mshatd}, we obtain
the following expression which relates the evolution of the interface
to the divergence of the cross-sectional flux of component A,
\begin{equation}\label{hmue}
 h_{t}+ q_{x}=0
\qquad \text{with }\quad q=\frac{1}{2}\,\frac{\partial}{\partial x} \int_0^{d} \mu_1 \, \mathrm{d}z.
\end{equation}
The last condition in \eqref{sc} together with \eqref{ffc} ensure
that the area of the film cross-section between the contact line and an
arbitrary but fixed cut-off $\hat x=L_\infty$ remains constant,
\begin{align}\label{mc}
\frac{d}{d{t}} \int_{ s({t})}^{L_\infty} h(x,{ t})\, \mathrm{d} x&=0,
\intertext{or, for $L_\infty\rightarrow\infty$,}
\frac{d}{d{t}} \int_{ s({t})}^\infty \left({ h( x,t)}-1\right)\, \mathrm{d} x&=\frac{d s}{d{t}}.
\label{mc2}
\end{align}

\subsection{Thin-film approximation}\label{sec:thin}

We can further approximate \eqref{mshat}, \eqref{sc}, \eqref{ffc} in
the limit of small contact angles, $\theta\ll 1$. We introduce the scalings 
\begin{subequations}
\label{eqn:thin_film_scalings}
\begin{align}
 x&=\frac{1}{\theta} \tilde x,&
 s&=\frac{1}{\theta}\tilde s, &
 t&=\frac{2}{\sigma\theta^4} \tilde t, \\
 \mu_1&=\frac{\sigma \theta^2}{2}\tilde \mu,&
 q&=\frac{\sigma \theta^3}{2}\tilde q,
\end{align}
\end{subequations}
and leave $z$ and $h$ unchanged. 
Inserting these we obtain 
\begin{subequations}
\label{msthin}
\begin{align}
  \theta^2 \timu_{\tix\tix} + \timu_{zz} &= 0,  \label{msthina}
\end{align}
in the domains $0<z<h(\tix,\tit)$ and $\tih(\tix,\tit)<z<d$, which is supplemented with the conditions
\begin{align}
\timu &= \frac{h_{\tix\tix}}{(1 + \theta^2h_{\tix}^2)^{3/2}}, \label{msthinb} \\
\theta^2 h_{\tit} &= \frac{1}{2}\left(\theta^2 [\timu_{\tix}]_{-}^{+}h_{\tix} - [\timu_{z}]_{-}^{+}\right),  \label{msthinc}
\end{align}
on the sharp interface $z=h(\tix,\tit)$ and 
\begin{align}
  \partial_z \timu &=0, \label{msthind}
\end{align}
on the substrates at $z = 0,\, d$.
\end{subequations}
The relation \eqref{hmue} remains unchanged
in the rescaled variables.
From the leading-order parts of \eqref{msthina}, \eqref{msthinb}, \eqref{msthind}, we immediately find $\timu=h_{\tix\tix}$, and
with the leading-order part of \eqref{hmue}, we obtain
\begin{subequations}\label{tfall}
\begin{equation}\label{tf}
h_\tit+\tiq_\tix=0, \qquad \text{where } \quad \tiq=\frac{d}{2}\,h_{\tix\tix\tix} .
\end{equation}
The leading-order contact line and far-field conditions are, respectively,
\begin{align}\label{thinbc1}
h&=0,\quad
h_\tix=1,\quad
\tiq=0, \quad \text{ at }  \tix=\tis,
\\ 
\label{thinbc2}
h &\rightarrow 1, \quad
\tiq \rightarrow 0 \quad \text{ at } \tix\rightarrow\infty.
\end{align}
\end{subequations}
From \eqref{mc2}, we get
\begin{equation}\label{lmc}
  \frac{d\tilde{s}}{d\tilde{t}}=\frac{d}{d\tilde{t}} \int_{\tilde{s}(\tilde{t})}^\infty \left( h(\tilde{x},\tilde{t})-1\right)\, \mathrm{d}\tilde{x}.
\end{equation}

\section{Topological transitions}\label{sec:res}

\subsection{Bilayer breakup}

The horizontal bilayer with an A-rich phase on top of a B-rich phase
that is created in the first step
is only metastable. The energy of the layer can be further decreased by
reducing the length of the interface between the two phases if the bilayers
are replaced by an arrangement of trapezoidal stripes. 
If, for example, we have a 50:50 ratio of the species, which in our
scalings implies a distance between the two substrates of $d=2$,
and we have neutral substrates, $\beta=0$, then the only contribution
to the energy comes from the interface between the phases. Moreover,
the stripes are rectangular in this case.  If the width of each stripe
is on average $w$, then a total interface length of $2n d$ for $2n$
stripes replaces a single interface of length $2n w$ for the bilayer
state.  Thus, the energy is reduced if and only if $w>d$.

\begin{figure}
\centering
\subfigure[]{\includegraphics[width=0.32\textwidth]{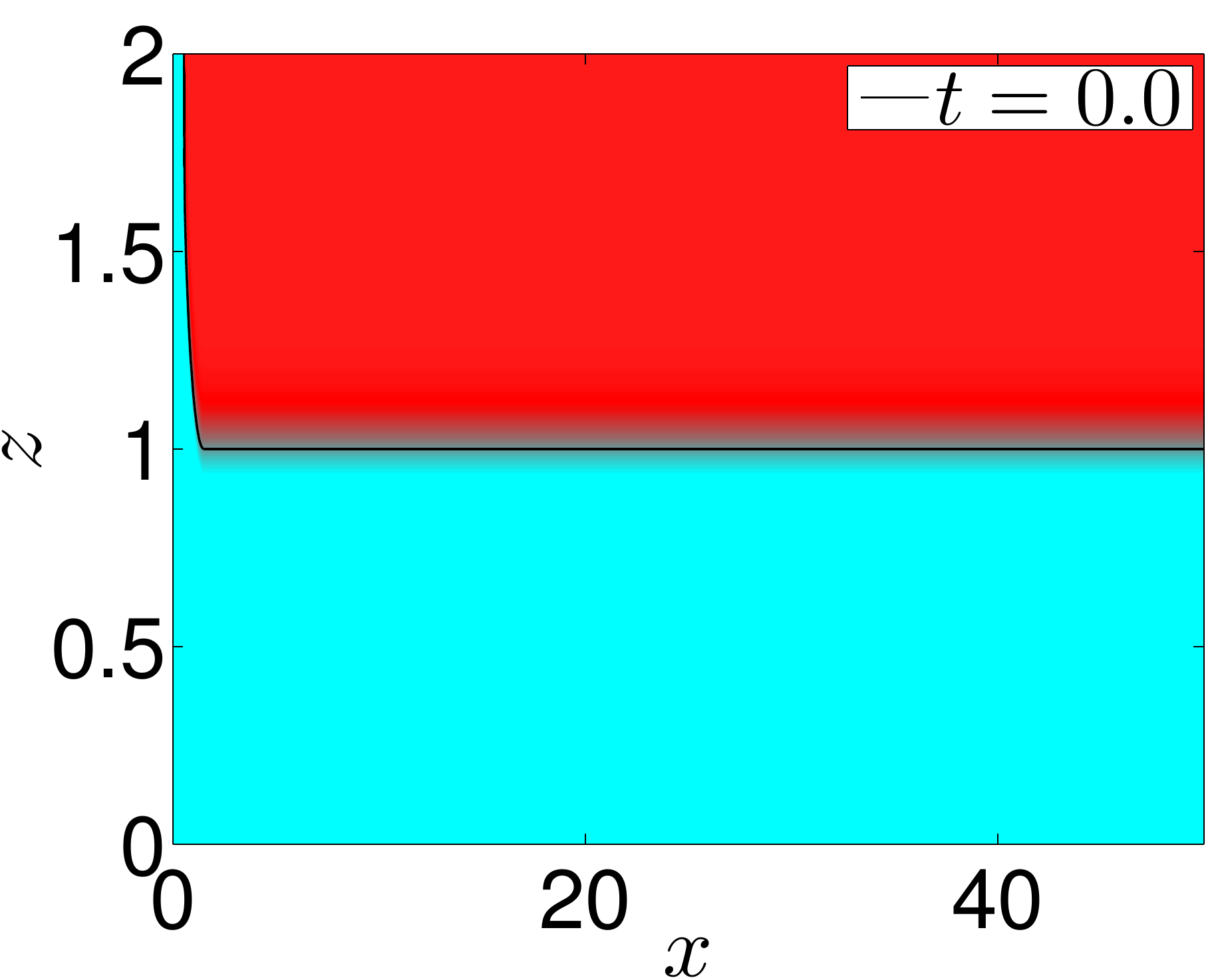}}
\subfigure[]{\includegraphics[width=0.32\textwidth]{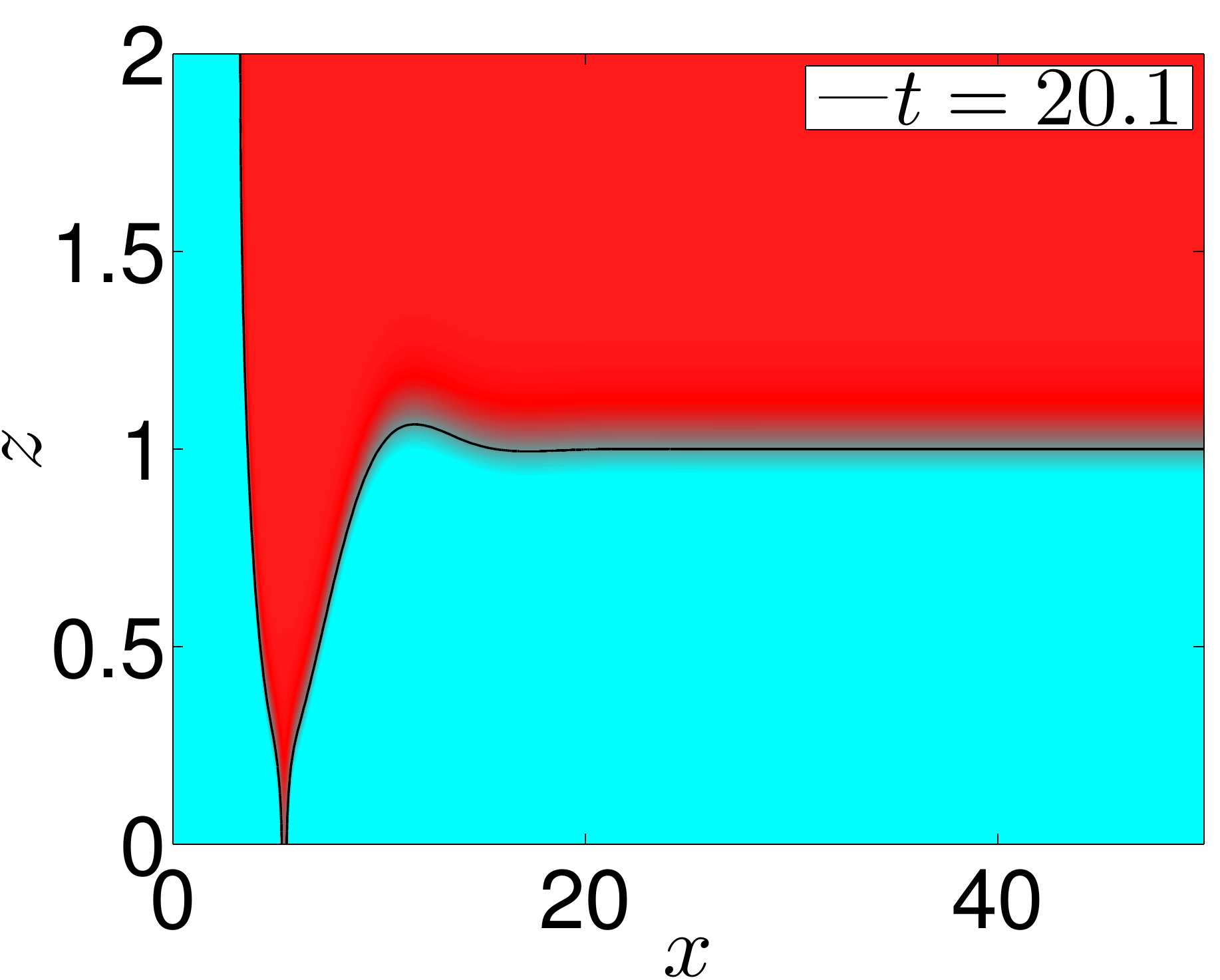}}
\subfigure[]{\includegraphics[width=0.32\textwidth]{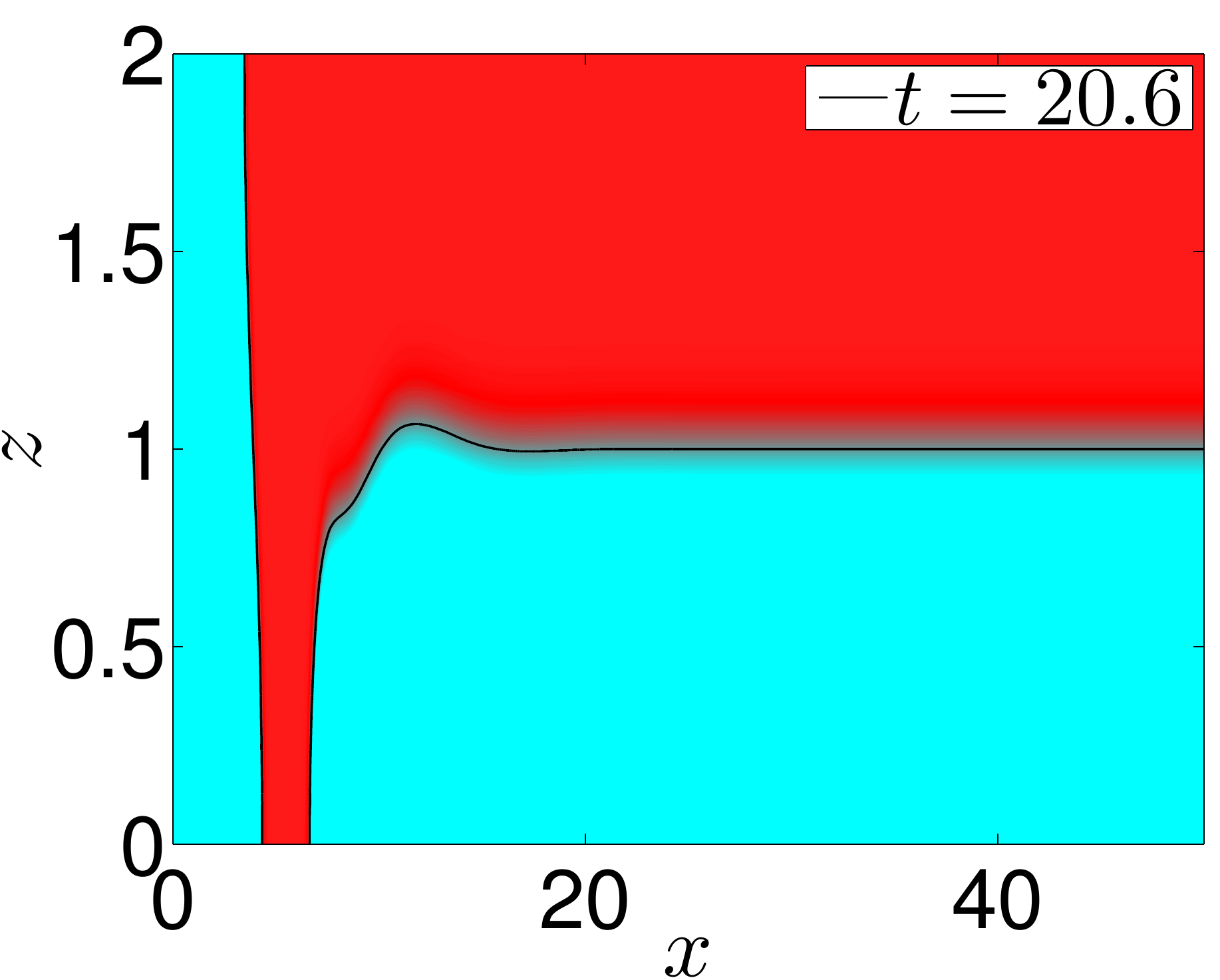}} \\
\subfigure[]{\includegraphics[width=0.32\textwidth]{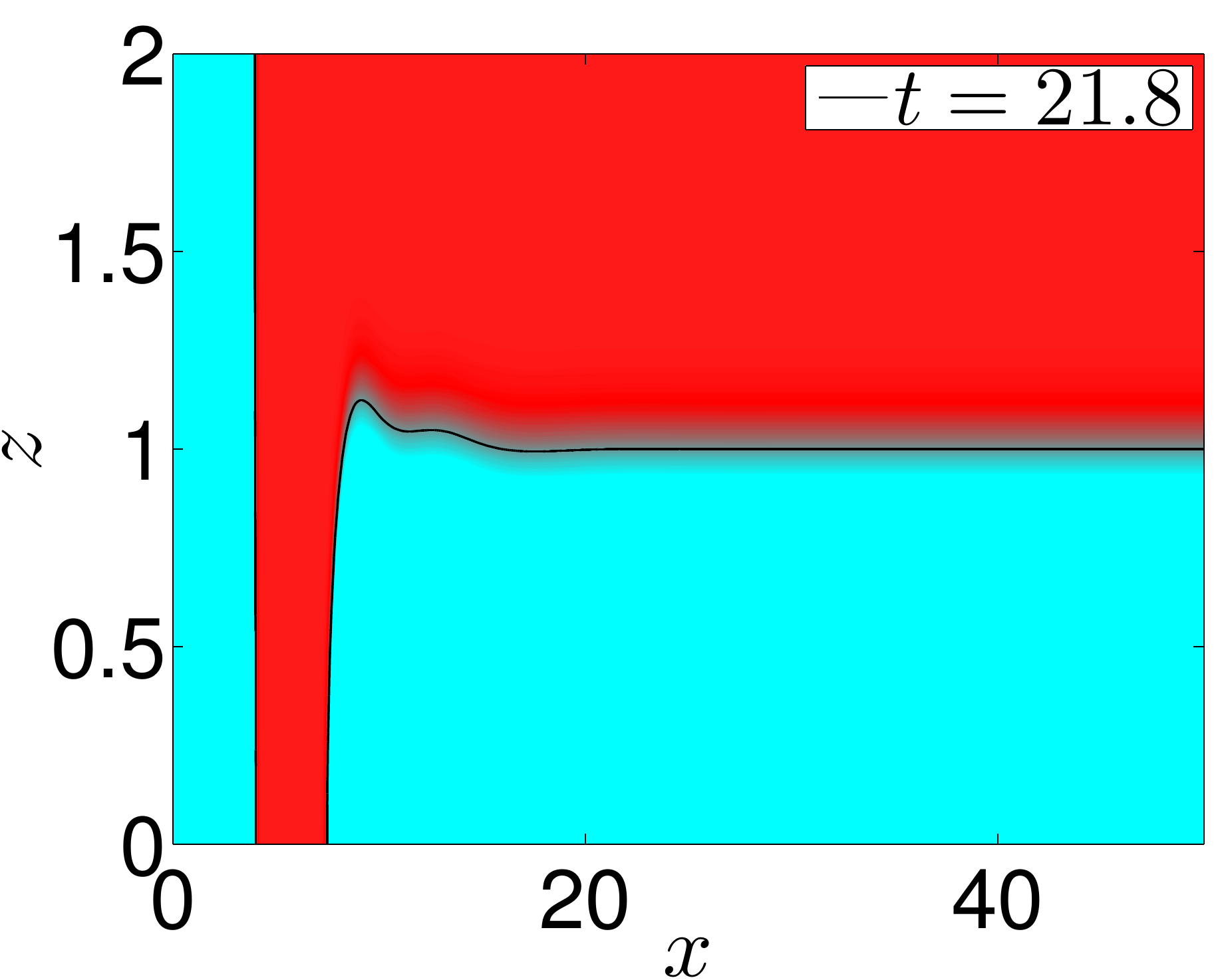}}
\subfigure[]{\includegraphics[width=0.32\textwidth]{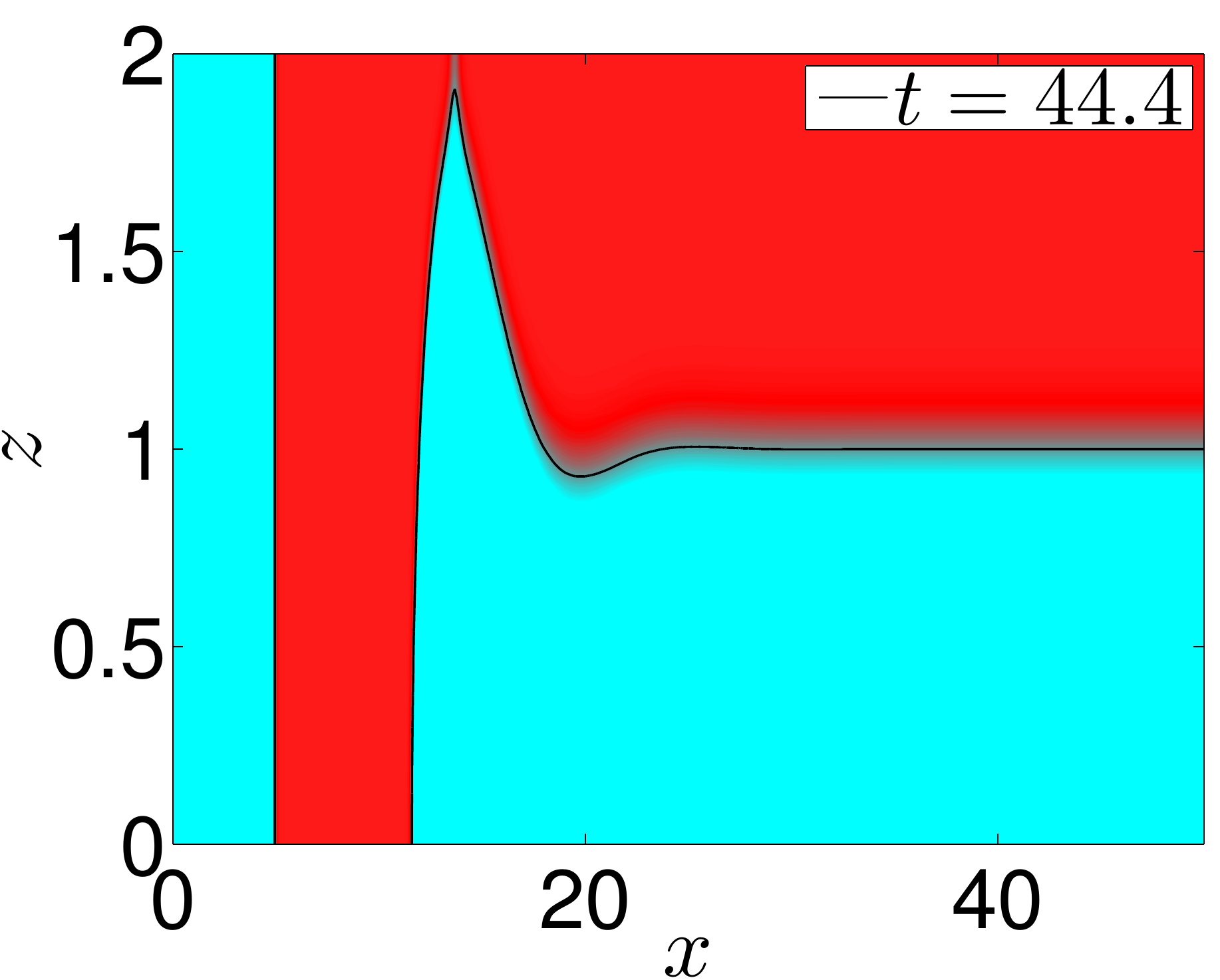}}
\subfigure[]{\includegraphics[width=0.32\textwidth]{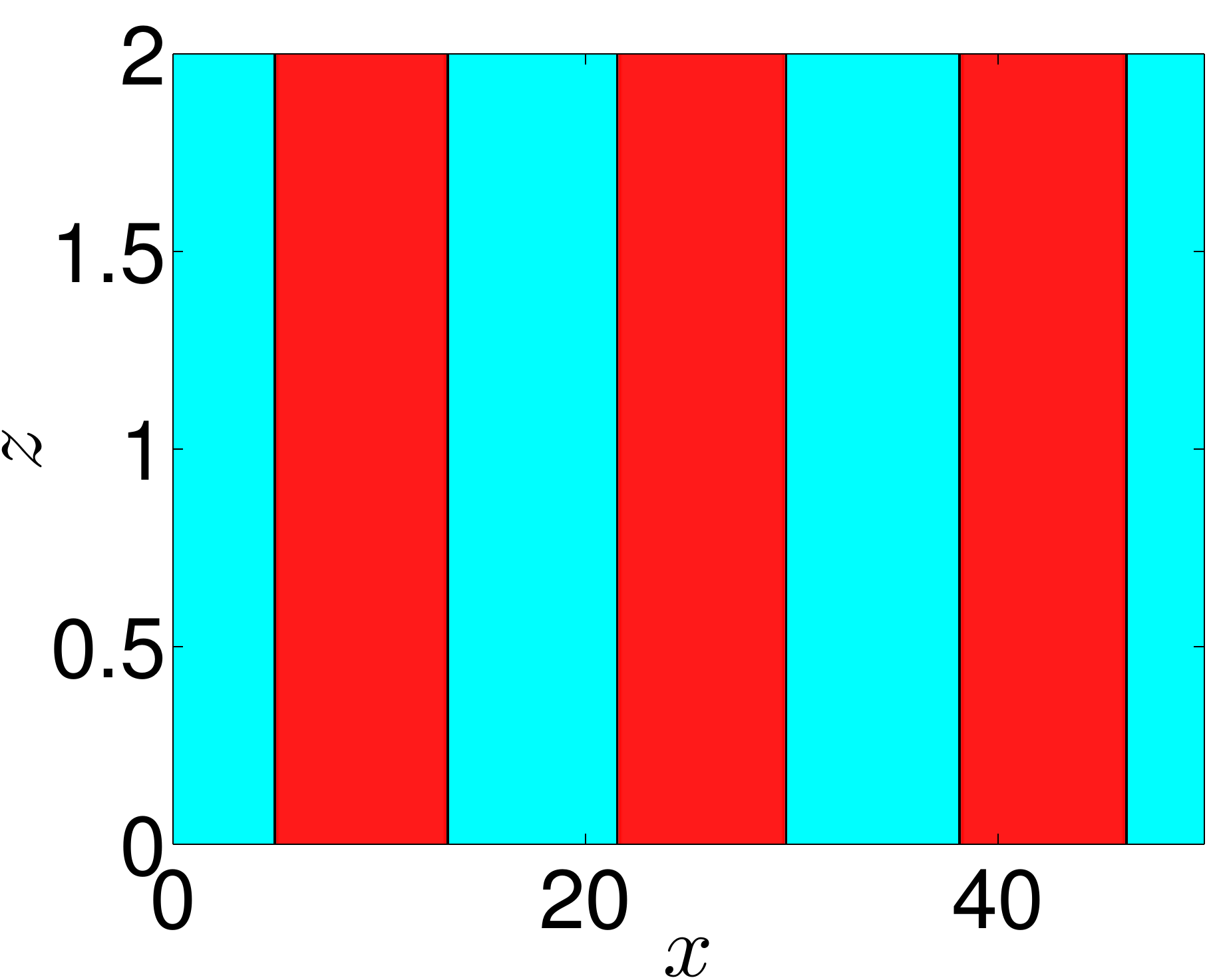}}
\subfigure[]{\includegraphics[clip=true, width=0.75\textwidth]{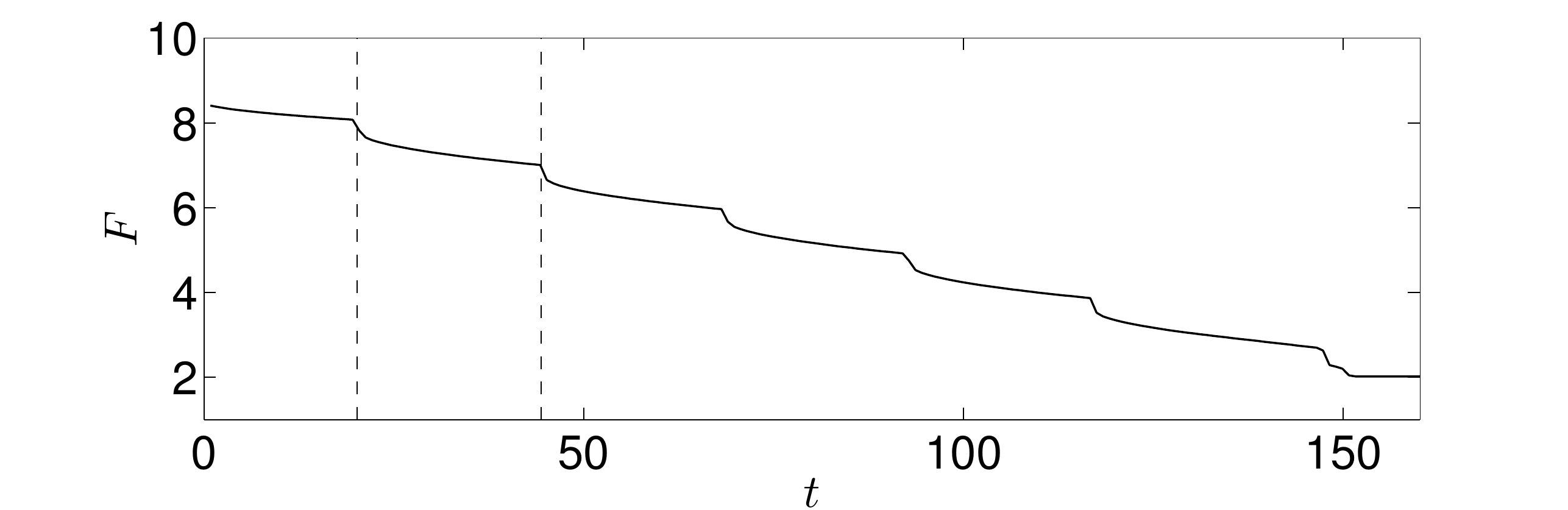}}
\caption{Top (a)--(c) and middle (d)--(f): The subsequent evolution of the solution
  in Fig.~\ref{fig:cooling}, showing, in particular, the dynamics that
  occur when a hole is introduced into the upper layer so that material from the
  bottom layer comes into contact with the upper substrate (a). The hole widens and the displaced
  mass creates a growing dip in the top layer that eventually touches the bottom substrate (b), creating
  new contact lines and a hole in the bottom layer. The new hole rapidly opens ((c) and (d)),
  and the material that is displaced in the bottom layer forms a growing ridge that eventually
  comes into contact with the upper substrate (e). The process then repeats itself until the bilayer
  has been tranformed into a sequence of columns.
  The solution is shown at times $t=0$, $t=20.2$, $t = 20.6$, $t = 21.8$, $t = 44.4$, and $t\to \infty$, i.e., the
  steady state. Bottom (g): The evolution of the free energy $F$. This can be proven to be 
  monotonic; see Appendix \ref{app:free_energy}. The sharp decreases correspond to creation of new contact lines
  and the rapid widening of the associated holes. The dashed lines
  correspond to the times shown in panels (b) and (e).}
\label{fig:nn}
\end{figure}

A refined energy argument, in combination with mass conservation,
has been given in \cite{Hennessy2013} for antisymmetric substrates
with general substrate-material interface energy densities (i.e.\ general $ \beta$).
This analysis reveal additional details about the transition between the two states if
the A-B interface is forced to touch one of the two substrates by
a finite perturbation when $\beta < 1$. As shown in Fig.~\ref{fig:nn}
the newly formed contact lines retract, to either
side, each collecting the A-rich phase in a growing trough. These
troughs eventually hit the bottom substrate and each gives rise to a new pair
of contact lines. The shedded A-rich material stabilises in a stripe,
while growing rims now appear in the B-rich phase until they hit the
upper substrate. The energy estimates show that the energy difference for
subsequent stripe formation is less than what is need for the formation of 
the first, so the process is self-sustaining: Once an initial ``hole'' is
formed, the entire bilayer will transform into an array of
stripes through a sequence of rupturing events. The argument also yields
estimates for the width of the stripes, which are $w=13.2/\theta$
for small contact angles and not more than $w=13.8/\theta$ for angles
up to $\pi/2$, measured at the center line.

We begin with the case where the substrate-material interface energy 
is weak and the equilibrium contact angle is large. 
Thus, we set the equilibrium contact angle
equal to $90^\circ$ which is equivalent to neglecting the 
energy of the substrate-material interfaces. It is assumed that a bilayer has formed, for example, by slowly
cooling the system, and we now investigate the dynamics of the system after this
bilayer has been ruptured. The corresponding initial condition for the phase-field
model is constructed with the leading-order solution of the
sharp-interface problem \eqref{eqn:si_comp_soln}
and taking the interface profile to be of the form
\begin{align}
  h(x) = \begin{cases} 2 - \sqrt{1 - (x - 1 - s_0)^2}, &\quad s_0 < x < s_0 + 1, \\
    1, &\quad x > s_0 + 1,
  \end{cases}
\end{align}
which represents a ruptured 50:50 bilayer with an initial contact line at $x = s_0$.
Setting $\theta = 90^\circ$ implies $\beta = 0$. We take $\varepsilon
= 0.127$ and $s_0 = 2/5$. The temperature is assumed to be constant so
we set $\chi\equiv 1$. The computational domain is cut off at $L_{\infty} = 50$. 

The results of a phase-field simulation are shown in Fig.~\ref{fig:nn}, and
it can be seen that puncturing the bilayer will induce a 
topological transition into a striped state. The stripes that form are 
perfect rectangles as a result of the equilibrium contact angle
being $90^\circ$, and the width of the second to fifth columns
are 8.4, 8.2, 8.2, and 8.4, respectively. The qualitative arguments
in \cite{Hennessy2013} yield an estimate of 8.8 for the stripe width,
which is in good agreement with the simulations. Also shown in this 
figure is the evolution of the free energy of the system, which for the
rescaled model in \eqref{beapp2} is defined at the end of Appendix \ref{app:free_energy}.
The free energy decreases monotonically, with large jumps occurring immediately after
new contact lines are created. For systems held at constant temperatures, the free
energy must be a monotonically decreasing function of time, and a proof of this claim
is given in Appendix \ref{app:free_energy}.

\begin{figure}
  \centering
  \subfigure[]{\includegraphics[clip=true,width=0.32\textwidth]{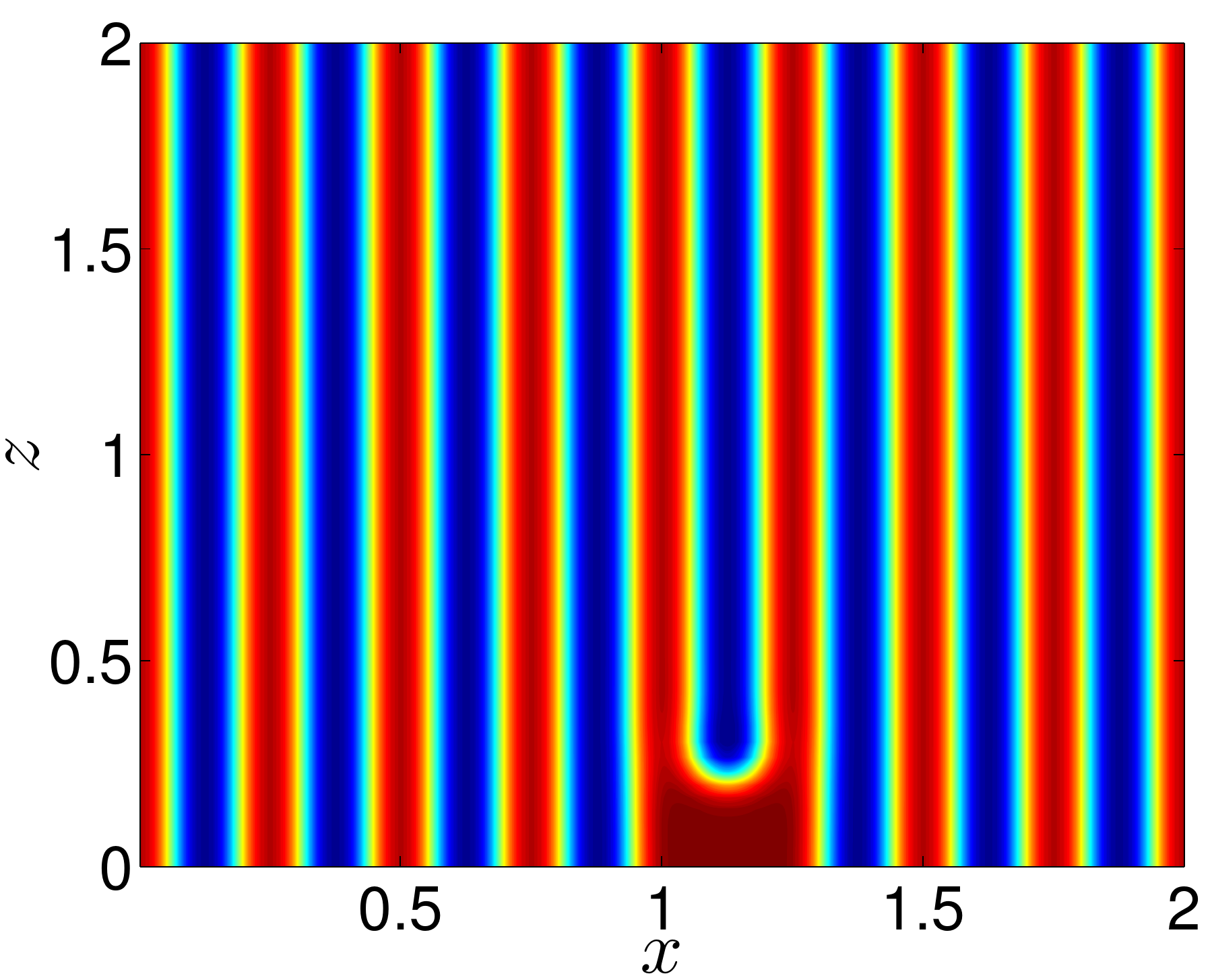}}
  \hspace{\fill}
  \subfigure[]{\includegraphics[clip=true,width=0.32\textwidth]{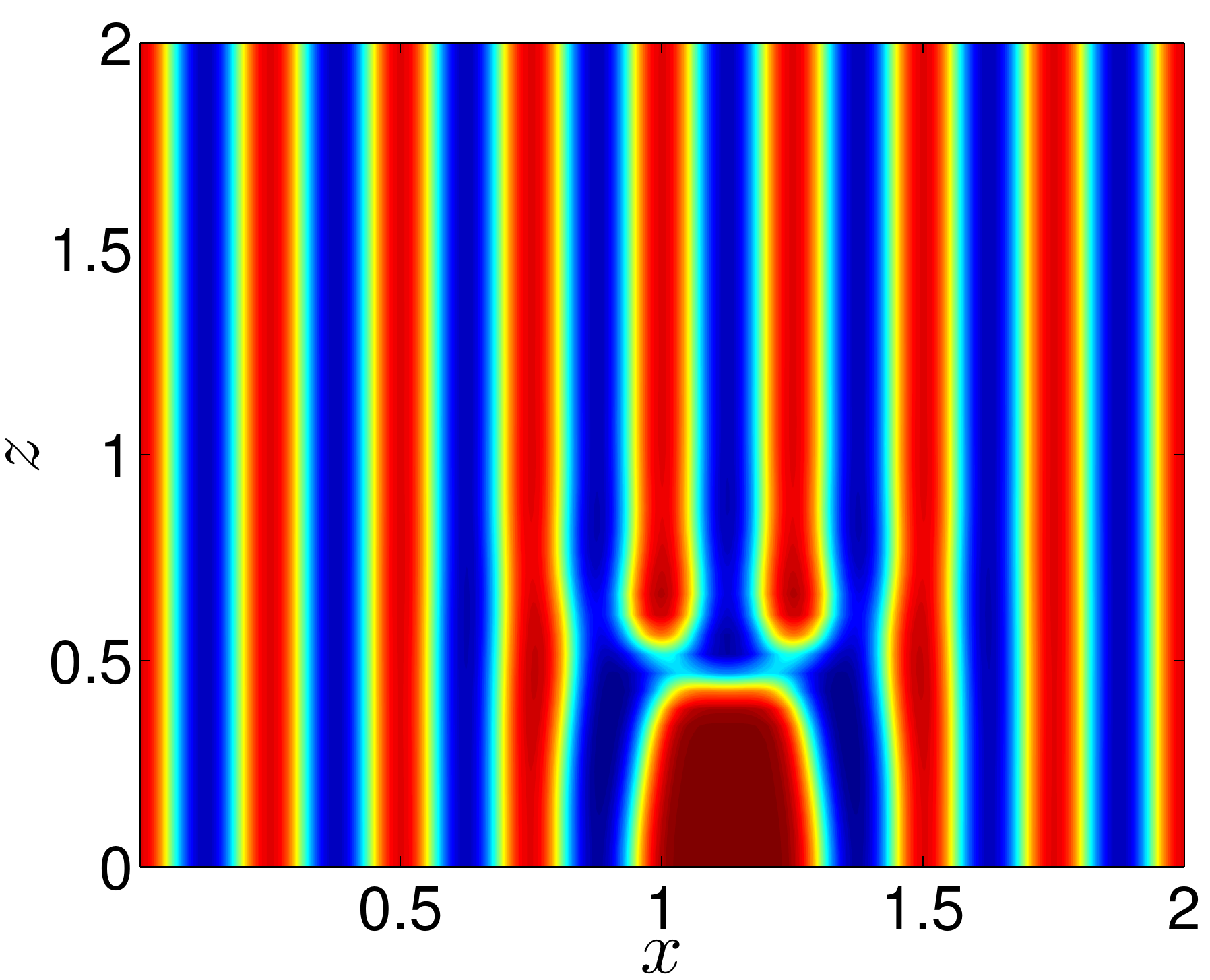}}
  \hspace{\fill}
  \subfigure[]{\includegraphics[clip=true,width=0.32\textwidth]{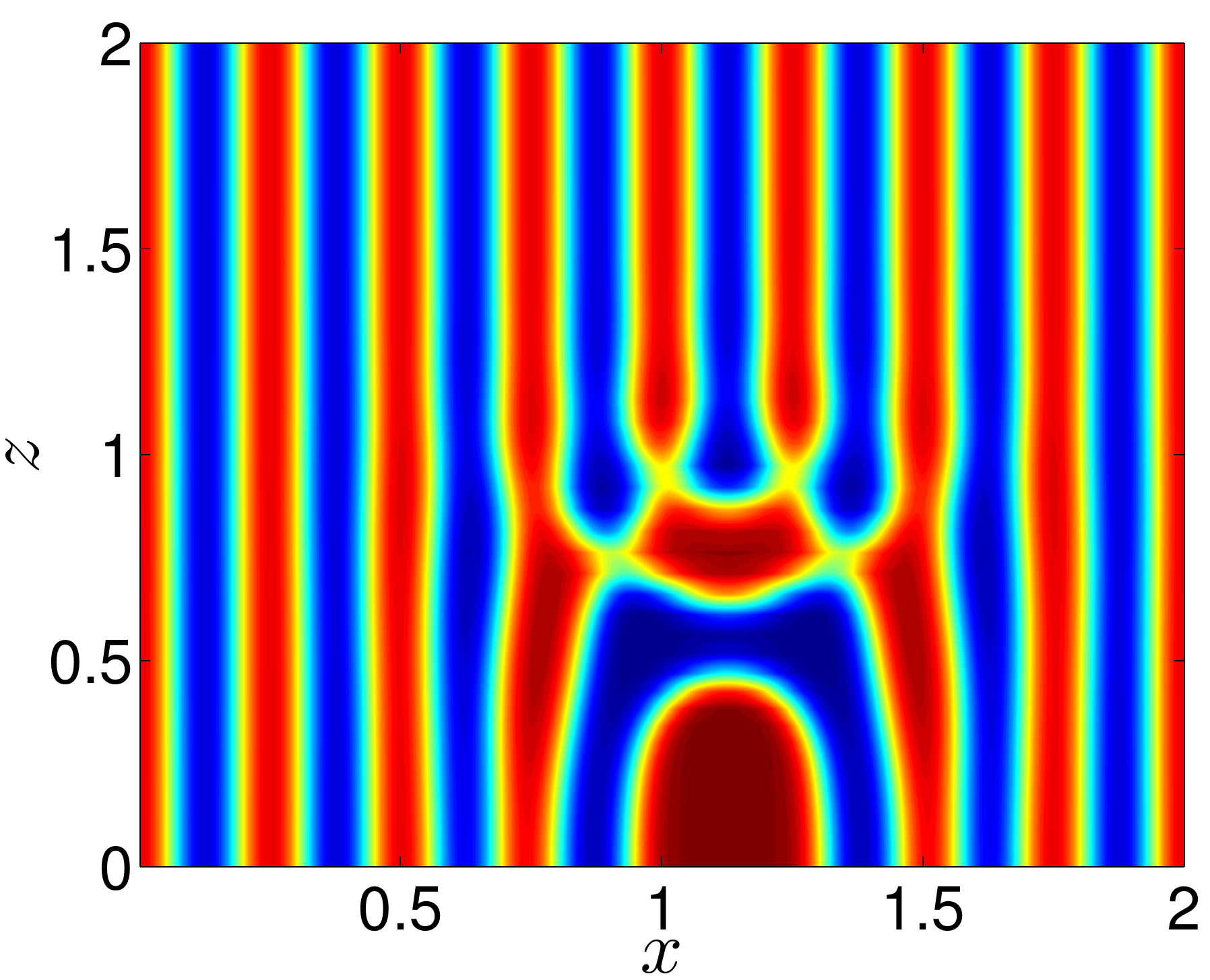}}
  \\
  \subfigure[]{\includegraphics[clip=true,width=0.32\textwidth]{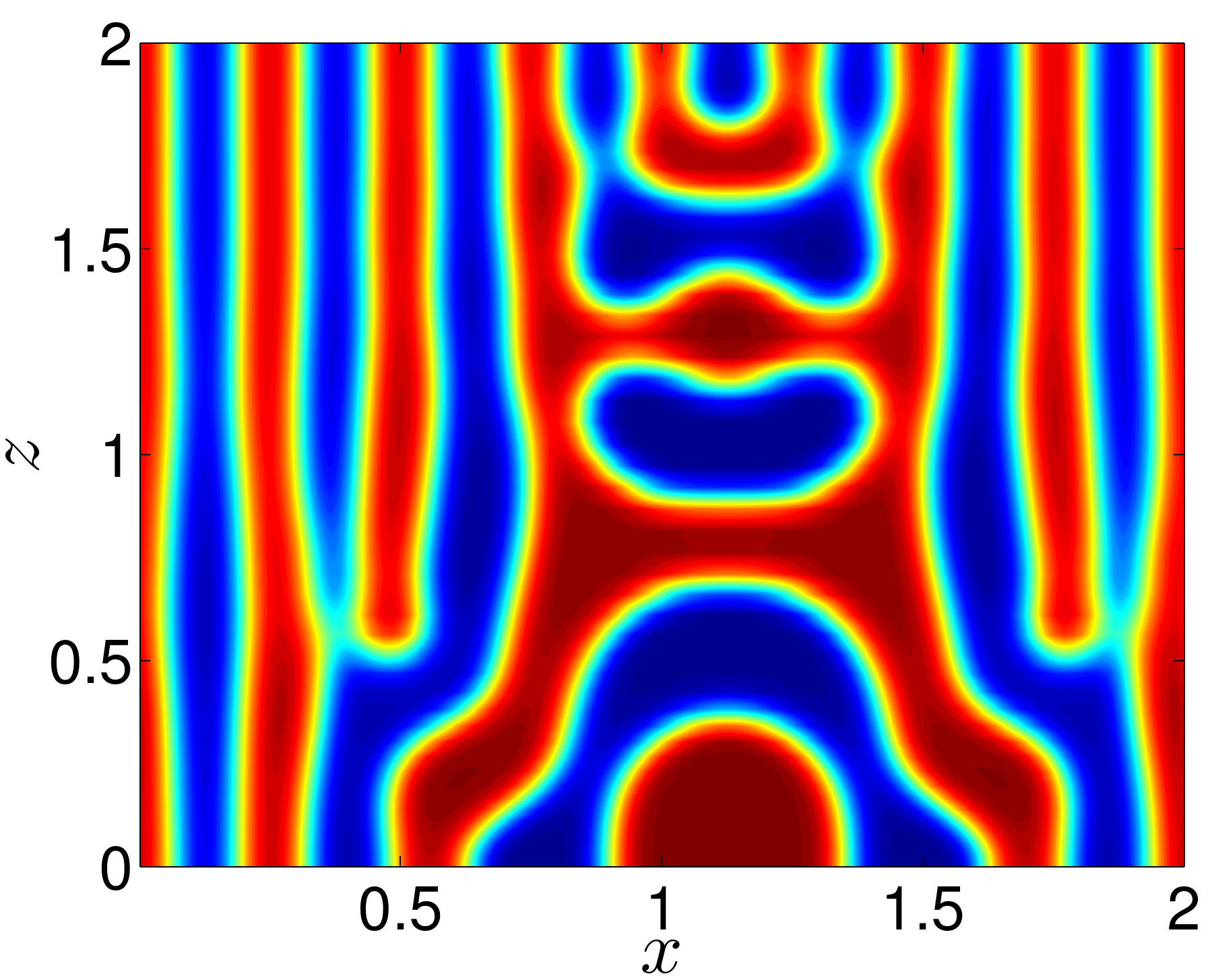}}
  \hspace{\fill}
  \subfigure[]{\includegraphics[clip=true,width=0.32\textwidth]{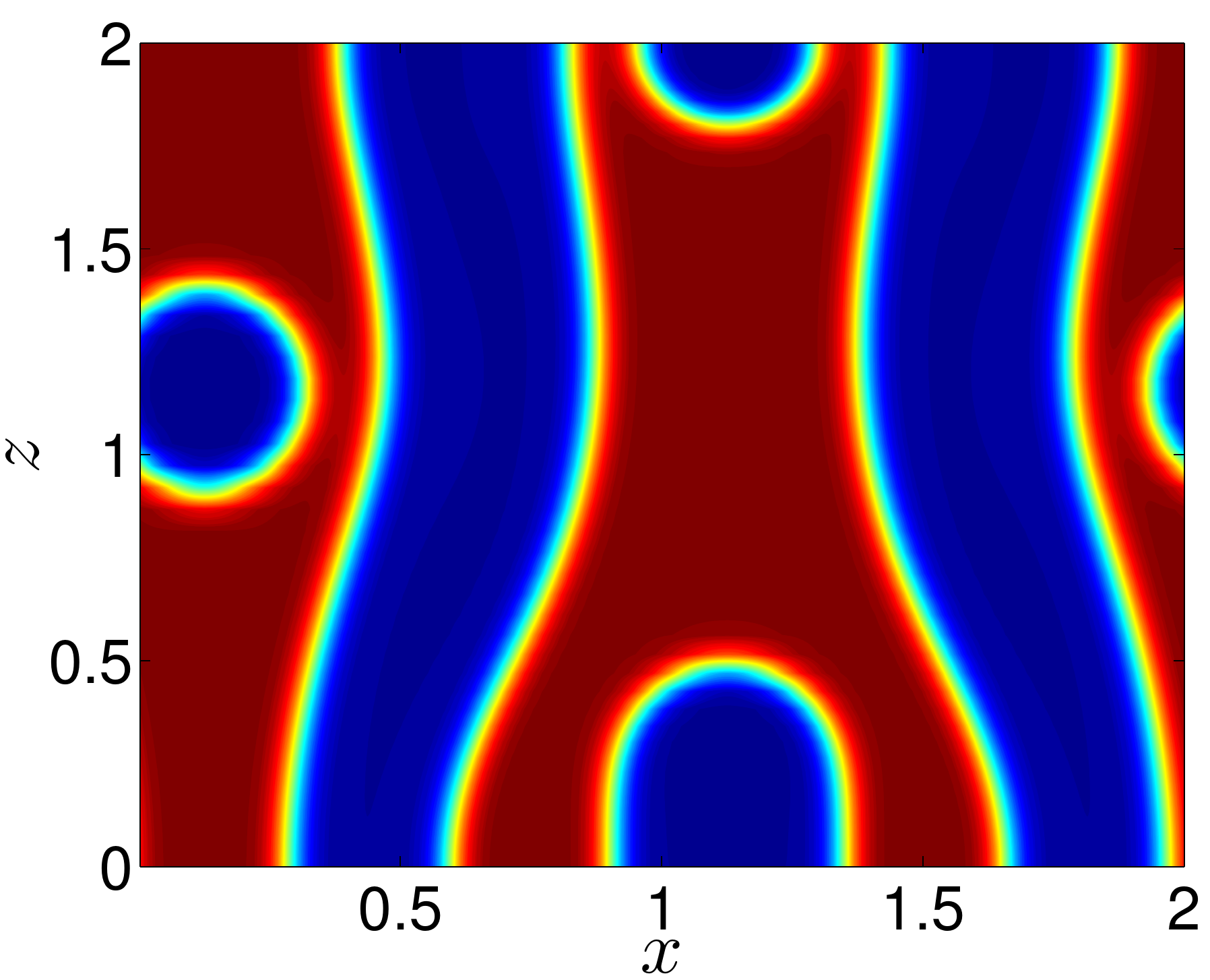}}
  \hspace{\fill}
  \subfigure[]{\includegraphics[clip=true,width=0.32\textwidth]{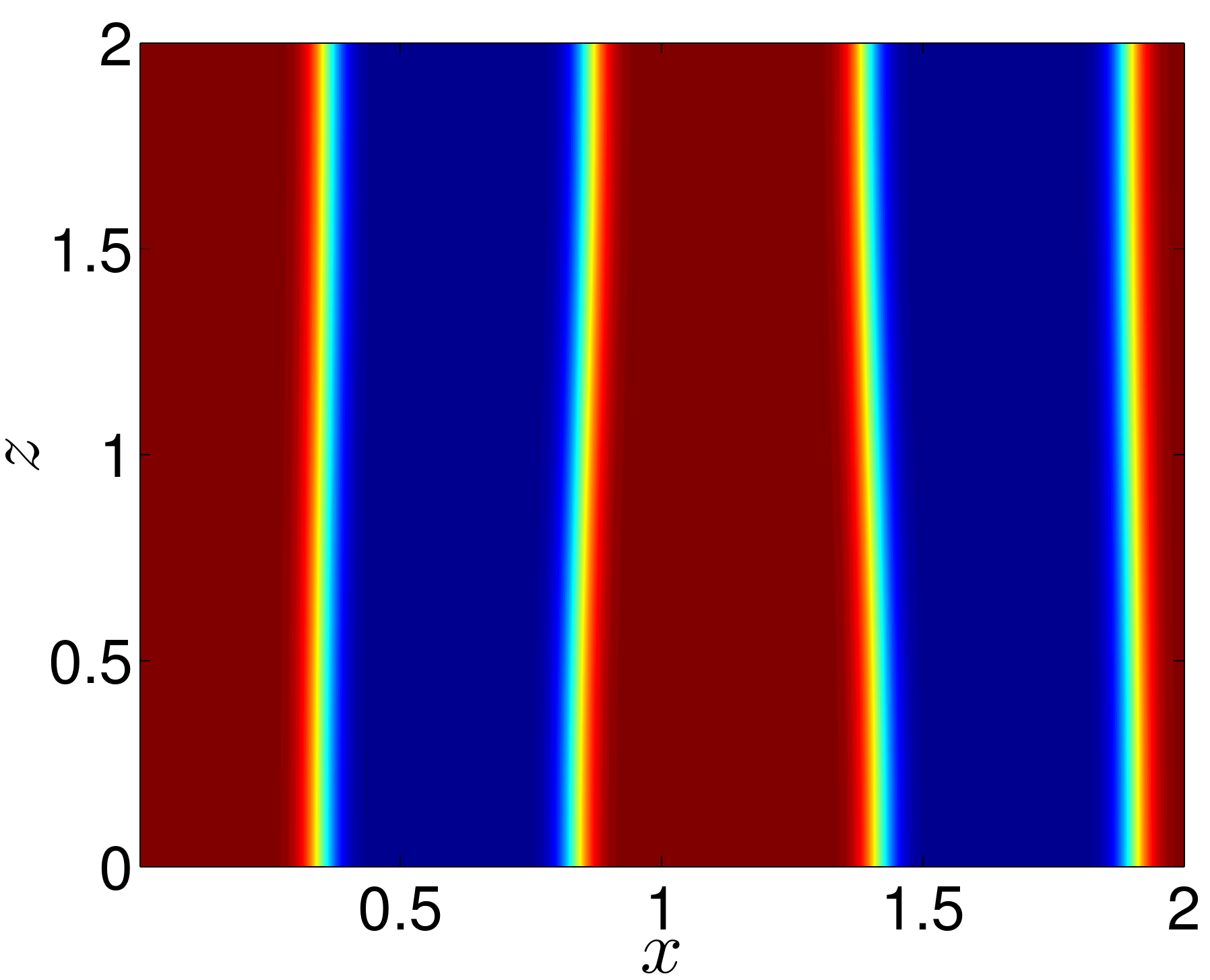}}
  \caption{Top (a)--(c) and bottom (d)--(f): Joining two columns of A-rich material
    initiates a rapid sequence of merging events that leads to a coarser
    set of columns. The columns of phase A are initially joined by creating a bridge between
    them at the bottom substrate, effectively shortening the column of B that separates them (a).
    The shortened column is pulled to the upper substrate to minimise its interface, but
    as this happens the two columns of A detach from the bridge (b). These also begin to 
    move upwards but as they do they come into contact with other columns of A (c)
    to initiate further detaching, merging, and coarsening events ((d) and (e)). The system then settles
    into a state of coarser columns (f). The solution is shown at times $t = 2.7 \times 10^{-5}$, $1.1\times 10^{-3}$, $2.1 \times 10^{-3}$, 
    $4.3 \times 10 ^{-3}$, $2.1 \times 10 ^{-2}$, and $0.37$.
    Merging two columns of the same phase in panel (f) causes the shortened
    column to retract; however, while it does it gets diffusively absorbed into the
    larger column of the same phase, resulting in a system with only one column
    of each phase.}
  \label{fig:reverse}
\end{figure}

\paragraph*{Reverse transformations}
Assume now that the mixture is separated into
$n$ pairs of adjacent A-B columns, much like the final configuration
shown in Fig.~\ref{fig:nn}. Using the same type of qualitative argument
that was presented in the previous paragraph, we find that such a configuration
will not be energetically favourable compared to the bilayer if the width of 
each column, $w$, is less than the height of the channel $d$, i.e., if $w < d$. 
In such a situation, we expect that a reverse topological transition will occur
if the initial state is perturbed. 

To explore this scenario, we perform simulations with the phase-field model
using initial conditions that correspond to a repeating sequence of A-B columns.
To perturb the system and initiate the transformation, we shorten one of the 
columns which has the effect of locally merging the two neighbouring columns. An
example of such an initial condition is shown in Fig.~\ref{fig:reverse} (a), 
where two columns of material A are brought together in system that begins as 8 pairs of
A-B columns. The other parameters values are $\varepsilon = 0.04$, $\chi \equiv 1$,
$\beta = 0$, with $L_\infty = d = 2$. Once the perturbation is added the 
system evolves under the action of interface minimisation and this drives
the shortened B column upwards and material A fills the void, thus thickening
the bridge that joins the two columns of A. However, around $t = 1.1\times 10^{-3}$
(Fig.~\ref{fig:reverse} (b)) the 
two A columns pinch off from the bridge and are pulled towards the upper substrate
in the same way that the shortened column of B was. Interactions between the
retracting A columns and their neighbours lead to additional merging and pinch-off
events (Fig.~\ref{fig:reverse} (c) and (d)), thus creating a very complex set of 
dynamics. For $t = 2.1\times 10^{-2}$ the morphology has been reduced to two pairs of deformed
A-B columns, with
the A columns containing pockets of material B. These pockets are diffusively absorbed into the
larger B columns (Fig.~\ref{fig:reverse} (e)), and eventually the system settles into a state
consisting of two pairs of A-B columns (Fig.~\ref{fig:reverse} (f)). This process can be
repeated in principle, leading to a fast coarsening of the striped morphology.

\subsubsection{Comparison to the thin-film model}
For our comparisons we choose again a 50:50 ratio of the two constituents but a shallow
contact angle so that the thin-film model \eqref{tfall} can be used
with $ d=2$.  We start from the bilayer situation where a layer
of the B-rich phase has formed at the bottom surface and a layer
of A-rich phase at the top, as a result, for example, of the slow
quenching process discussed Section~\ref{sec:bil}.  

For the numerical solution of the thin-film model, we truncate the
domain at $\tilde x=L_\infty$ (with a choice for $L_\infty$ that was
larger than 100) and impose $ h_{\tilde x}=1$, $h_{\tilde x\tilde x\tilde x}=0$ 
there.  At $\tilde x=\tilde s$, we
impose $h=0$ and $h_{\tilde x}=1$ and require global mass
conservation to hold,
\[
\int_{\tilde s(\tilde t)}^{L_\infty} h(\tilde x,\tilde t)\,\mathrm{d}\tilde x
=\int_{\tilde s(0)}^{L_\infty} h(\tilde x,0)\,\mathrm{d}\tilde x.
\] 
The truncated domain is mapped onto the unit interval by the linear
transformation $\tilde x\mapsto (\tilde x-\tilde s(\tilde t))/(L_\infty-\tilde
s(\tilde t))$ and the resulting problem is discretised using finite
differences in space and implicit Euler in time. Step
doubling was used to control the time discretisation error.
Initial conditions at $\tilde t=0$ are
\begin{subequations}\label{ic}
\begin{align}\label{ica}
\tilde s&=0, \qquad
h(\tilde x,0)= d- h_i(\tilde x), &\\
\intertext{where}
h_i(\tilde x)&=
\begin{cases}
\tilde x-\tilde x^2/4  & \text{ for } \tilde x< 2,\\
1& \text{ elsewhere. } 
\end{cases}
\label{icb}
\end{align}
\end{subequations}
Notice that with this choice, we assume that initially, 
a thin hole filled by B-rich phase has been created in the A-rich top layer 
giving rise to a contact line at the top substrate, $z=d=2$. 
The initial interface $z=h(\tilde x,0)$  
satisfies the contact angle condition $ h_{\tilde x}=-1$ at
$\tilde x=\tilde s$.  
  
\begin{figure}
\begin{center}
\includegraphics[clip=true,width=0.32\textwidth]{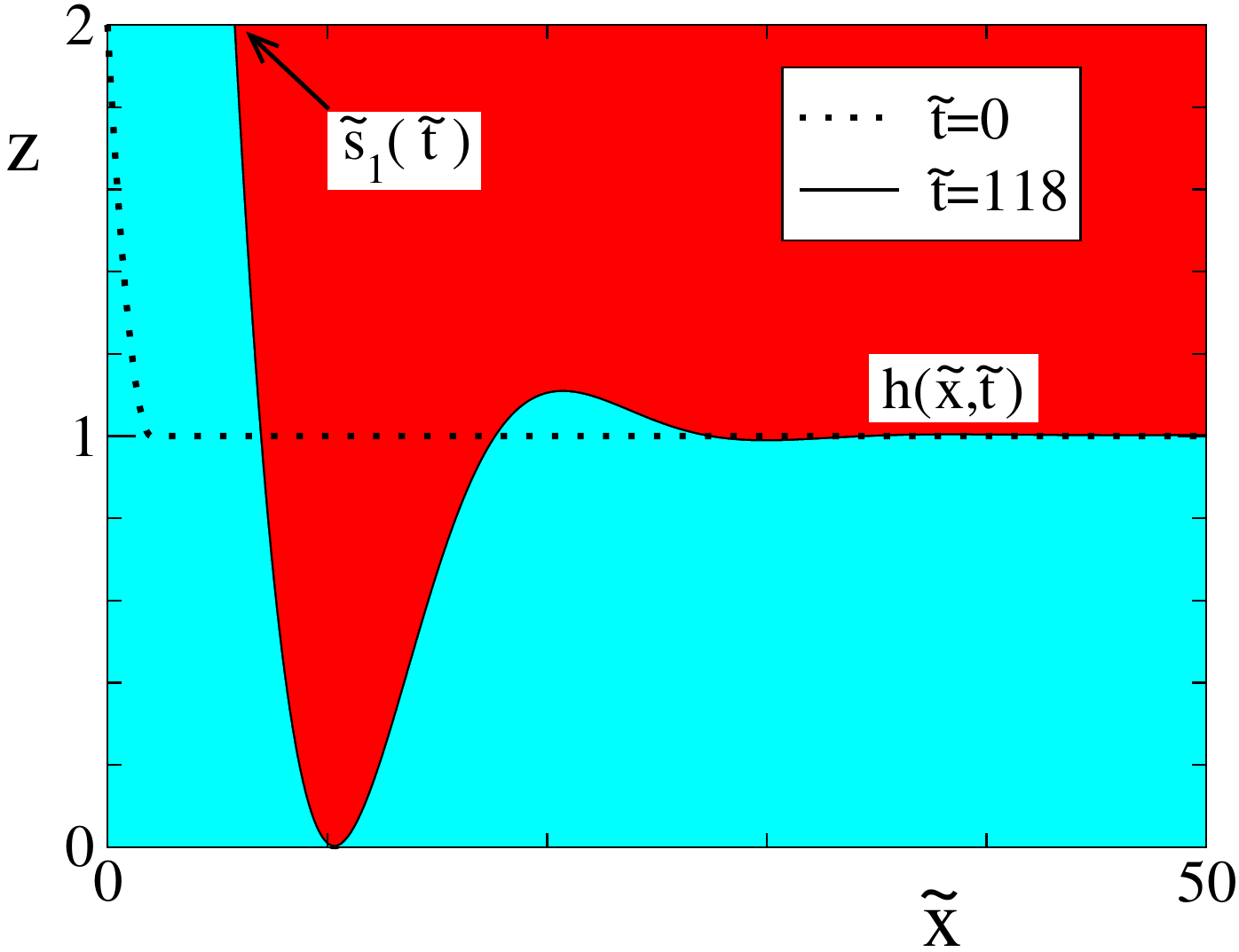}
\hspace{\fill}
\includegraphics[clip=true,width=0.32\textwidth]{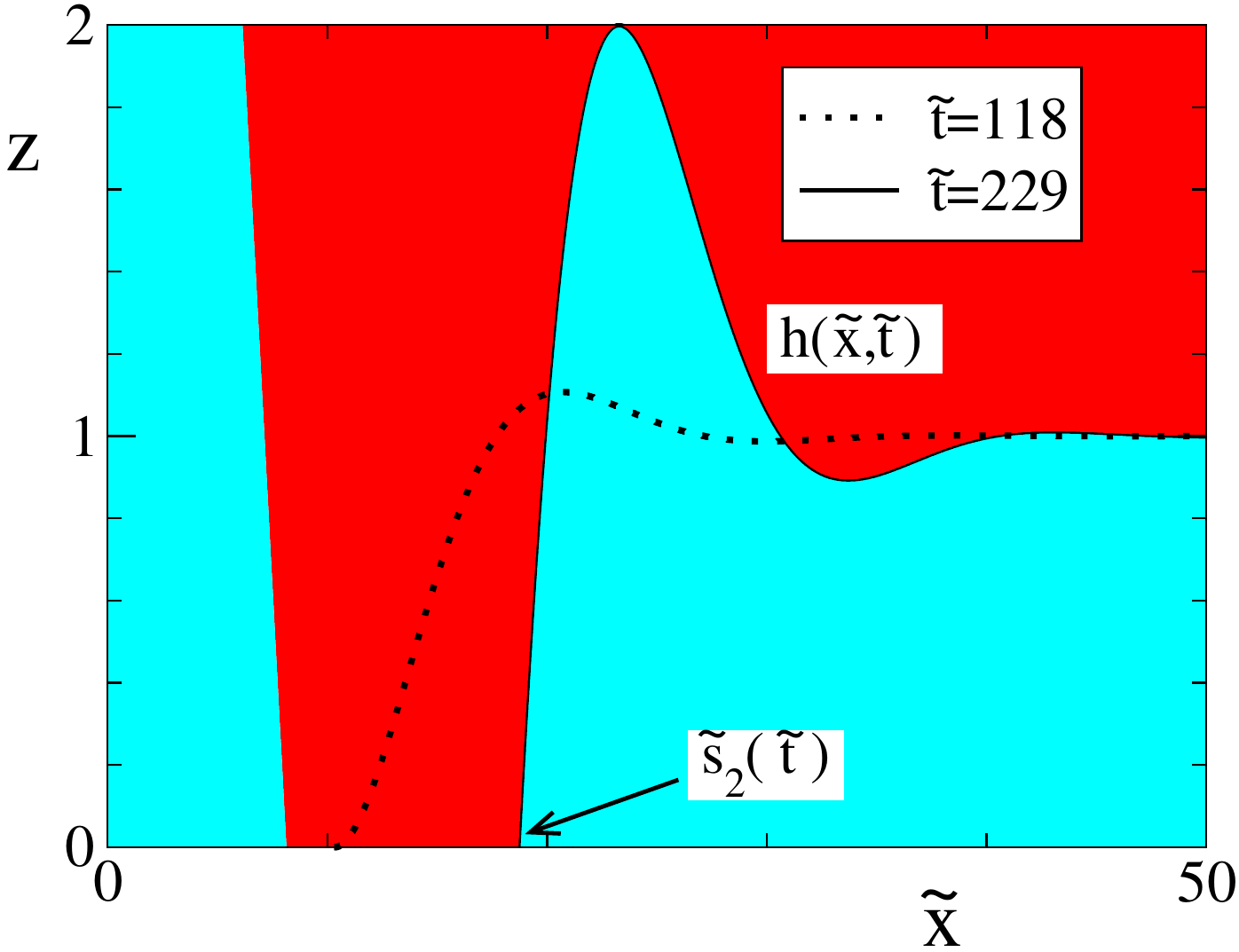}
\hspace{\fill}
\includegraphics[clip=true,width=0.32\textwidth]{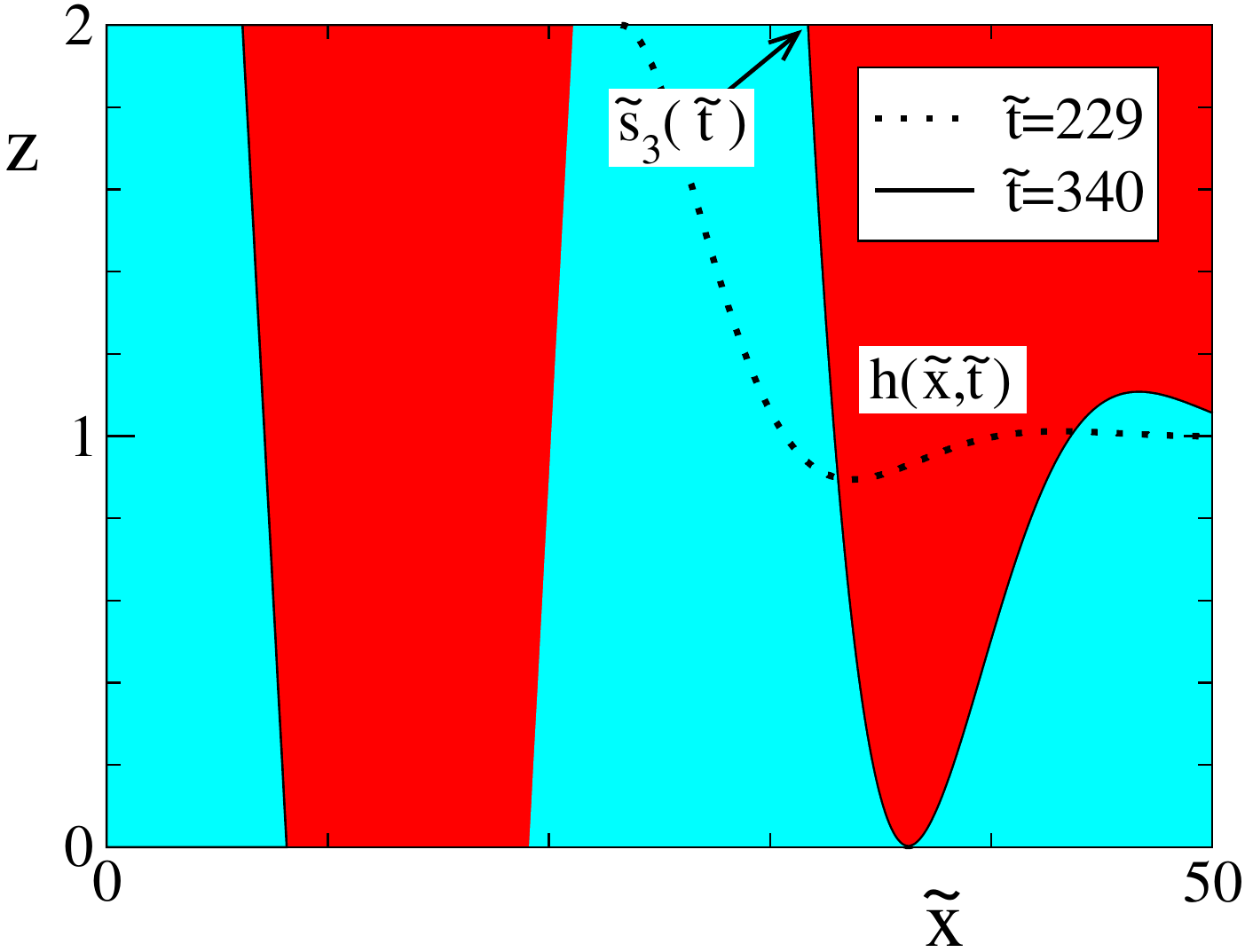}
\\
\includegraphics[clip=true,width=0.32\textwidth]{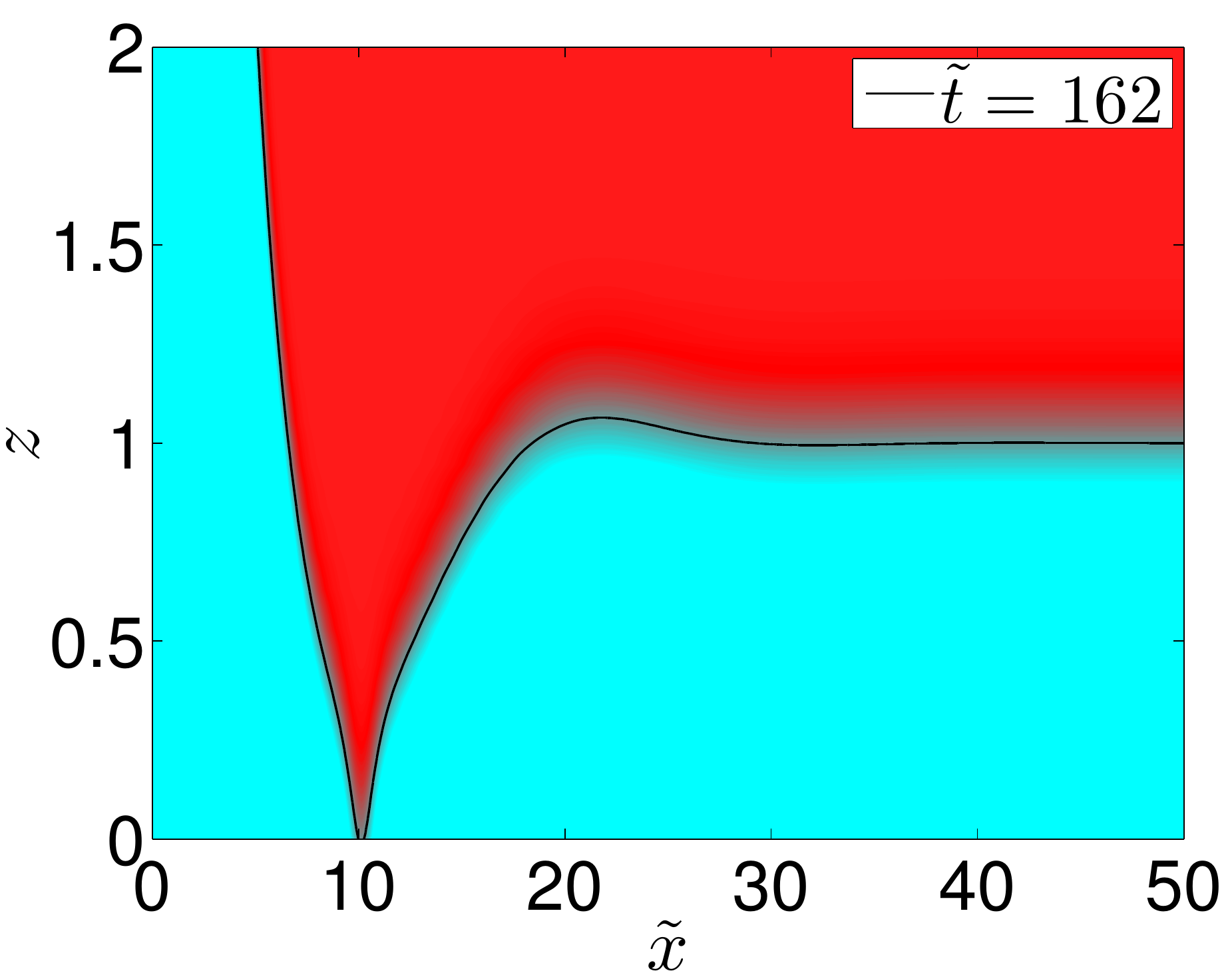}
\hspace{\fill}
\includegraphics[clip=true,width=0.32\textwidth]{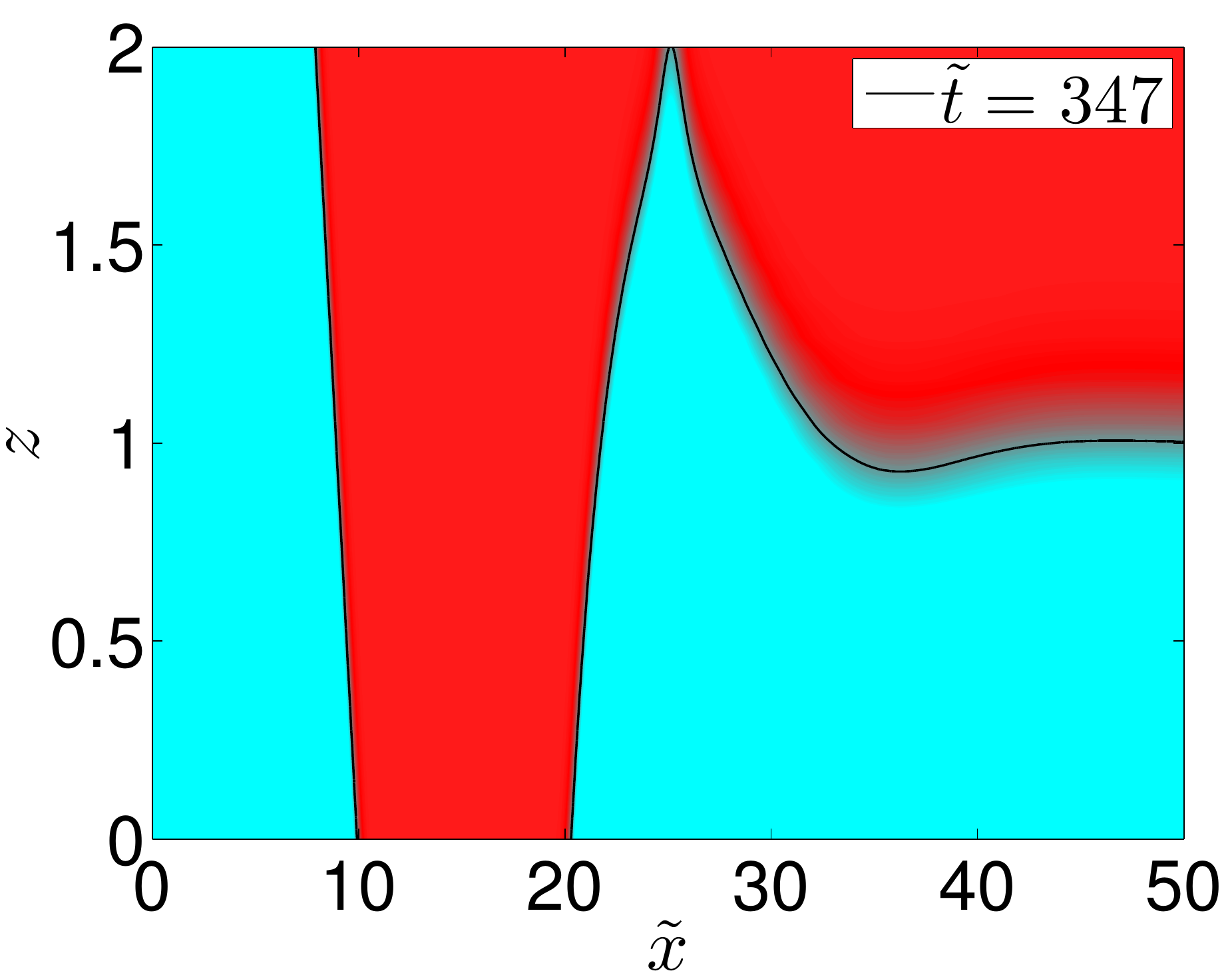}
\hspace{\fill}
\includegraphics[clip=true,width=0.32\textwidth]{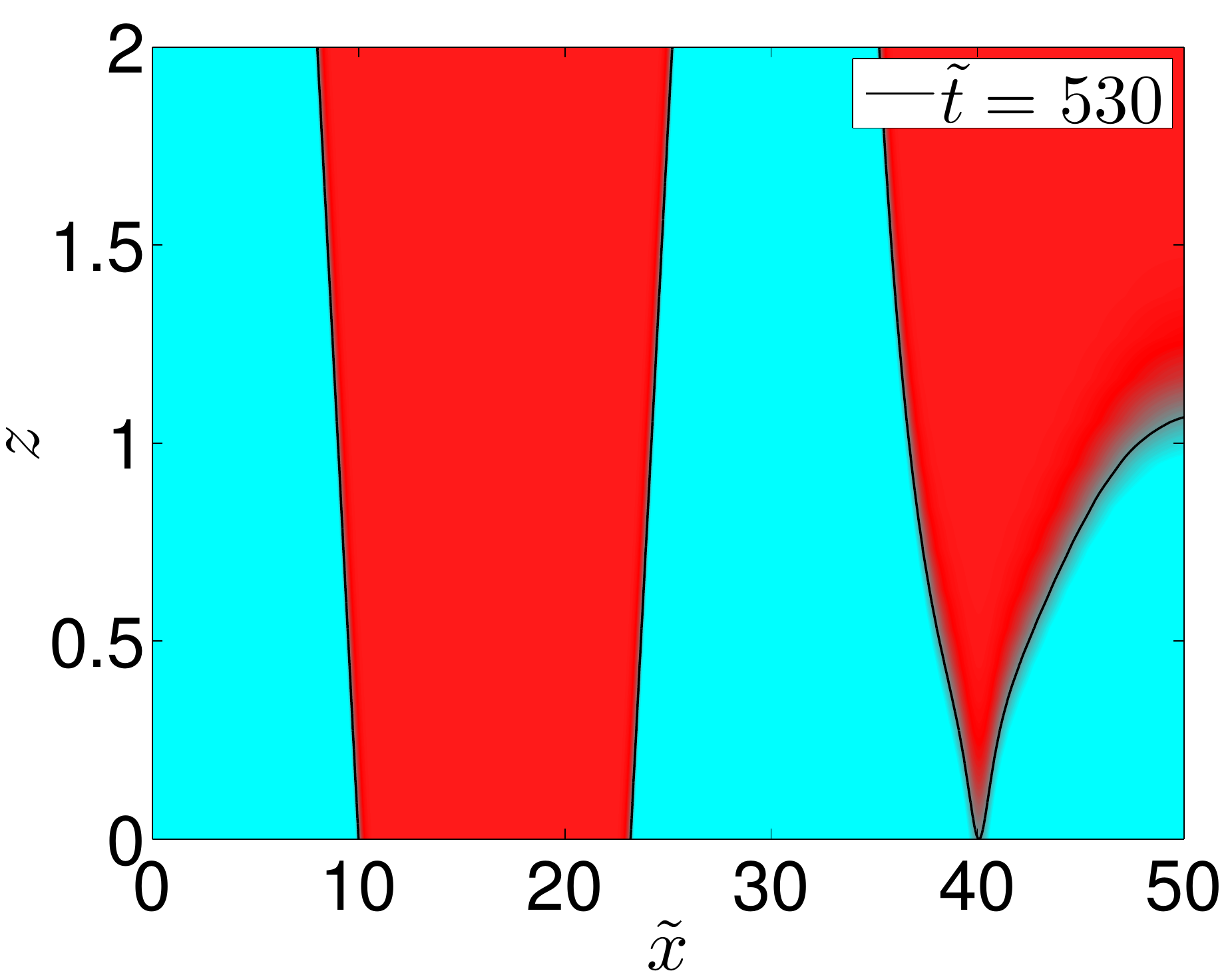}
\end{center}
\caption{\label{fig:double}%
Evolution of the interface between the phases after introduction of
an initial hole with B-rich phase (blue) 
penetrating the A-rich phase (red) at $\tilde x=0$, $\tilde z=2$.  
The top row (with the subfigures labelled (a)-(c))
and the bottom row ((d)-(f)) show the numerical results for the 
thin-film sharp-interface model \eqref{tfall}  
and for the phase-field
model \eqref{beapp2}
respectively.  The initial hole widens and pushes the B-phase
to the right. This entails growth of the minima and maxima in the
interface which gives rise to alternating holes in the A- and B-layer
and thus the formation of new contact lines. 
Further details are given in the text.}
\end{figure}

In Fig.~\ref{fig:double}(a), the initial data ($\tilde t=0$) is shown
by a dotted line: a hole filled with B-rich material has been created in the A-rich phase.
The contact line at 
$x=\tilde s_1(0)=0$ retreats, until
the minimum $\min_{x} h$  of the interface  hits $ z=0$
at time $\tilde t_1=118$ and position $\tilde x_{2}$ and forms a
pair of new contact lines, one of which moves to the left, the other,
labelled $\tilde s_2(\tilde t)$, to the right.  Thus, the B-rich layer is
split into two parts.  The left part settles into an equilibrium,
shown in Fig.~\ref{fig:double}(b), with the interface between the
the B- and A-rich phase located at $z= 8.14-\tilde{x}$ as determined
by conservation of the B-rich phase in the leftmost stripe.

The contact line $ z=0$, $\tilde x =
\tilde s_2(\tilde t)$ for the other part travels to the right, with
a growing rim forming in the B-layer ahead of it.  The interface
$h(\tilde x,\tilde t)$ for $\tilde x<\tilde s_2(\tilde t)$ is
obtained by restarting the simulations at $\tilde t=\tilde t_1$, with
the initial position for the contact line at $\tilde s_2=\tilde x_2$
and using $h(\tilde x,\tilde t_1)$, for $x>\tilde s_2(\tilde
t_1)$ as initial profile (indicated by a dotted line in (b)).

Eventually, in Fig.~\ref{fig:double}(b), the right moving and growing
ridge hits the top substrate at $\tilde t_3=229$ and $\tilde x=x_3$,
giving rise to another pair of contact lines.  This splits the
A-rich layer into two parts. The left part equilibrates as a strip of A-rich phase
between the old B-A interface, at $ z=-\tilde x+8.14$, and the new A-B
interface at $z=\tilde x-19.1$. The last expression follows from
conservation of phase A.  The contact line at $z=2$, $\tilde x =
\tilde s_3(\tilde t)$ for right part of the A-rich layer continues 
to evolve, with a decreasing minimum ahead of it, that will
eventually result in another rupture of the B-rich layer shown in 
Fig~\ref{fig:double}(c). 
The evolution up to this point is obtained
by restarting the simulation at $\tilde t=\tilde t_3$ with
$\tilde s_{3}(\tilde t_2)=x_2$, using $h(\tilde x,\tilde t_2)$
for $\tilde x>\tilde s_3(\tilde t_3)$ as initial interface profile.
As before, this initial profile is indicated in the figure by a
dotted line. The width of the first equilibrated A-rich stripe is
13.0, which is very close to the prediction of 13.2 from the qualitative argument
made in the beginning of this section and in \cite{Hennessy2013}. 

The accuracy of the thin-film model can be examined by
running equivalent numerical simulations using the phase-field
model in \eqref{beapp2} and applying the thin-film scalings in
\eqref{eqn:thin_film_scalings} to the results. The thin-film initial
condition given in \eqref{ic} is converted into an initial
condition for the phase-field model using the leading-order composite
solution for the sharp-interface model \eqref{eqn:si_comp_soln} after 
the appropriate rescalings have been made. 


When applying the thin-film scalings a value for the equilibrium
contact angle $\theta$ is needed.
We choose a value of $\theta = 50^\circ$, which is
large enough to allow the topological transition to be computed in a
reasonable amount of time but small enough that the thin-film
limit is still captured in the phase-field model. We also
replace $\theta$ by $\tan\theta$ in the thin-film scalings \eqref{eqn:thin_film_scalings} to account
for the loss of accuracy in the small-angle approximation.

The bottom row of Fig.~\ref{fig:double} shows the results of a
phase-field simulation in the thin-film limit. 
We have also taken $\chi \equiv 1$ and $\varepsilon = 0.2$. 
A contact angle of $50^\circ$ corresponds
to a substrate-material interface energy given by $\beta = 0.64$. The computational
domain is truncated at $x = 80$ which, in the thin-film scaling,
corresponds to $L_\infty = 95$. 

The simulation
confirms that an initial hole at the top of the bilayer can lead to
a receding contact line, which, in turn, will create a growing ridge
in the bottom layer. We find that at time $\tilde{t} = 162$ this ridge
comes into contact with the bottom substrate and new contact lines are
born at position $\tilde{x} = 8.7$ (see Fig.~\ref{fig:double}(d)). The
motion of the new contact line on the bottom substrate creates a ridge in
the upper layer which comes into contact with the top substrate at time
$\tilde{t} = 347$ (shown in Fig.~\ref{fig:double}(e)). The two contact
lines on the bottom substrate settle into their equilibrium positions at
at $\tix = 8.6$ and $\tilde{x} = 21.9$. Moreover, the new contact line on
the upper substrate creates another ridge in the lower layer which comes
into contact with the bottom substrate at time $\tilde{t} = 530$. This is
shown in Fig.~\ref{fig:double}(f). 

We have continued the phase-field simulation
until the topological transition is complete and the bilayer has
been transformed into a series of trapezoidal columns. The widths
of the second to fifth columns as measured from the line $z = 1$
are given by $15.3$, $15.0$, $15.0$, and $14.8$, respectively. These
values are in good agreement with the predicted value of 13.5 obtained 
by qualitative arguments. 

\begin{figure}
  \centering
  \includegraphics[clip=true,width=0.9\textwidth]{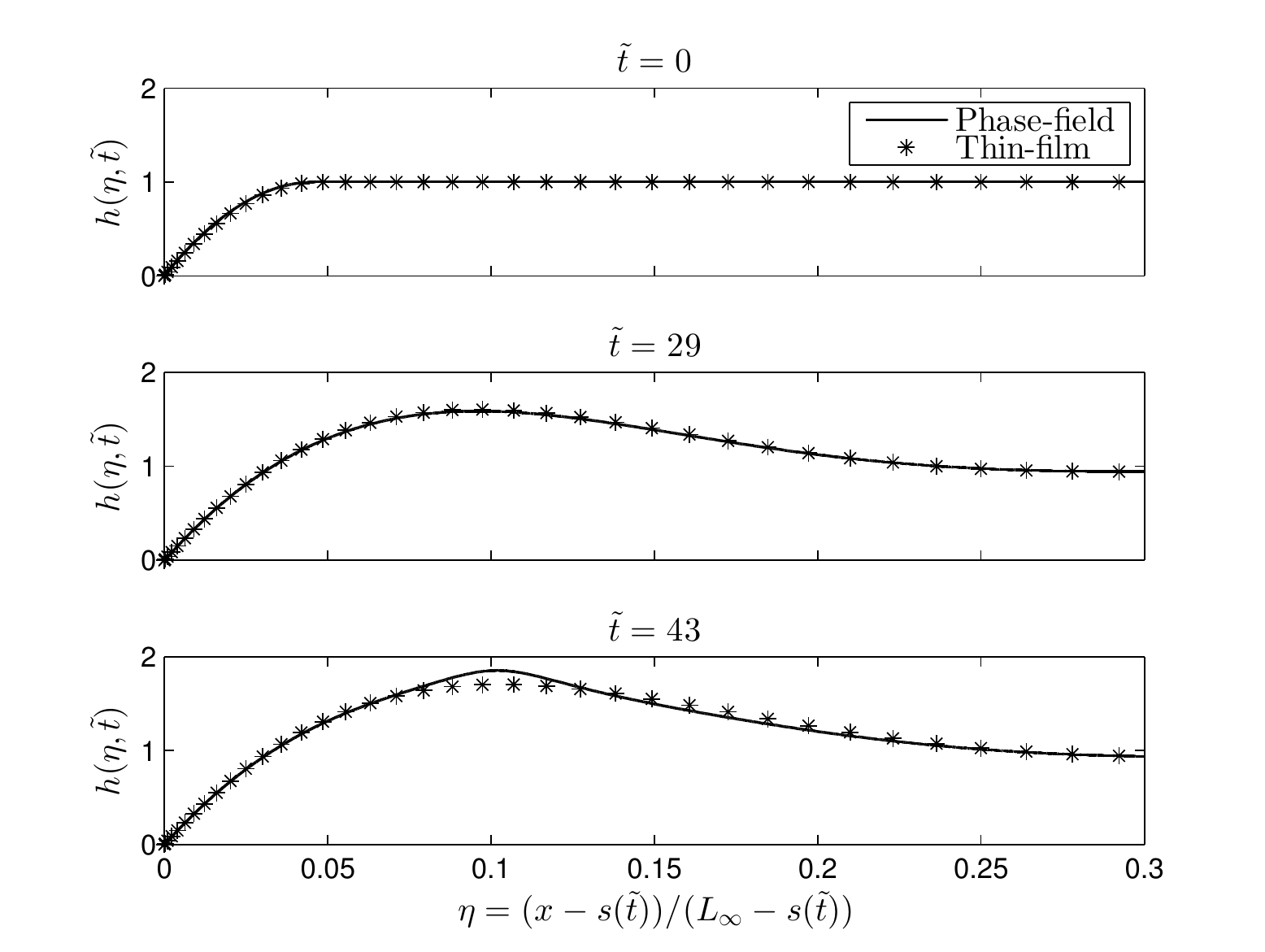}
  \caption{Comparison of the bilayer interface profiles computed using the phase-field
    and thin-film models when the equilibrium contact angle is set to $\theta
    = 20^\circ$. Before the system is close to a topological transition ($\tilde{t} \leq 29$),
    the agreement between the two models is excellent. However, when the maximum
    of the ridge is within an $O(\varepsilon)$ distance from the upper substrate ($\tilde{t} = 43$),
    a ``suction'' effect pulls the interface upwards in the phase-field model, making the
    transition occur sooner than in the thin-film model where this effect is absent.}
  \label{fig:theta_20}
\end{figure}

From these results we conclude that the thin-film and phase-field
models agree remarkably well on the geometrical aspects of the 
topological transition. However, when comparing when each transition
occurs in the two models, we see that there are significant quantitative
discrepancies. To test whether these differences are a consequence of using
a large contact angle in the phase-field model which might prevent the 
thin-film regime from being accurately captured, the first topological transition
has been computed with the equilibrium contact angle reduced to $20^\circ$. 
All of the other parameters are kept the same as above. 
Fig.~\ref{fig:theta_20} compares the interface profiles computed
using the phase-field and thin-film models at various times. The agreement between the 
models is excellent for $\tilde{t} \leq 29$; however, 
differences in the solutions exist for larger times. The source
of this discrepancy is due to an apparent ``suction'' effect that occurs in 
the phase-field model when the interfacial ridge gets within an $O(\varepsilon)$
distance from the upper substrate. This effectively pulls the ridge up towards
the substrate, making the topological transition occur sooner in the phase-field
model than in the thin-film model, where this suction effect is absent. Thus,
the discrepancy between the times of the topological transitions is not
necessarily due
to the phase-field model being outside of the thin-film regime, but rather it is
caused by  substrate-interface interactions that are neglected in the sharp-interface and thin-film
models.

\paragraph*{Competition between layer thickness and rim shedding}

\begin{figure}
\centering
\subfigure[]{\includegraphics[clip=true,width=0.5\textwidth]{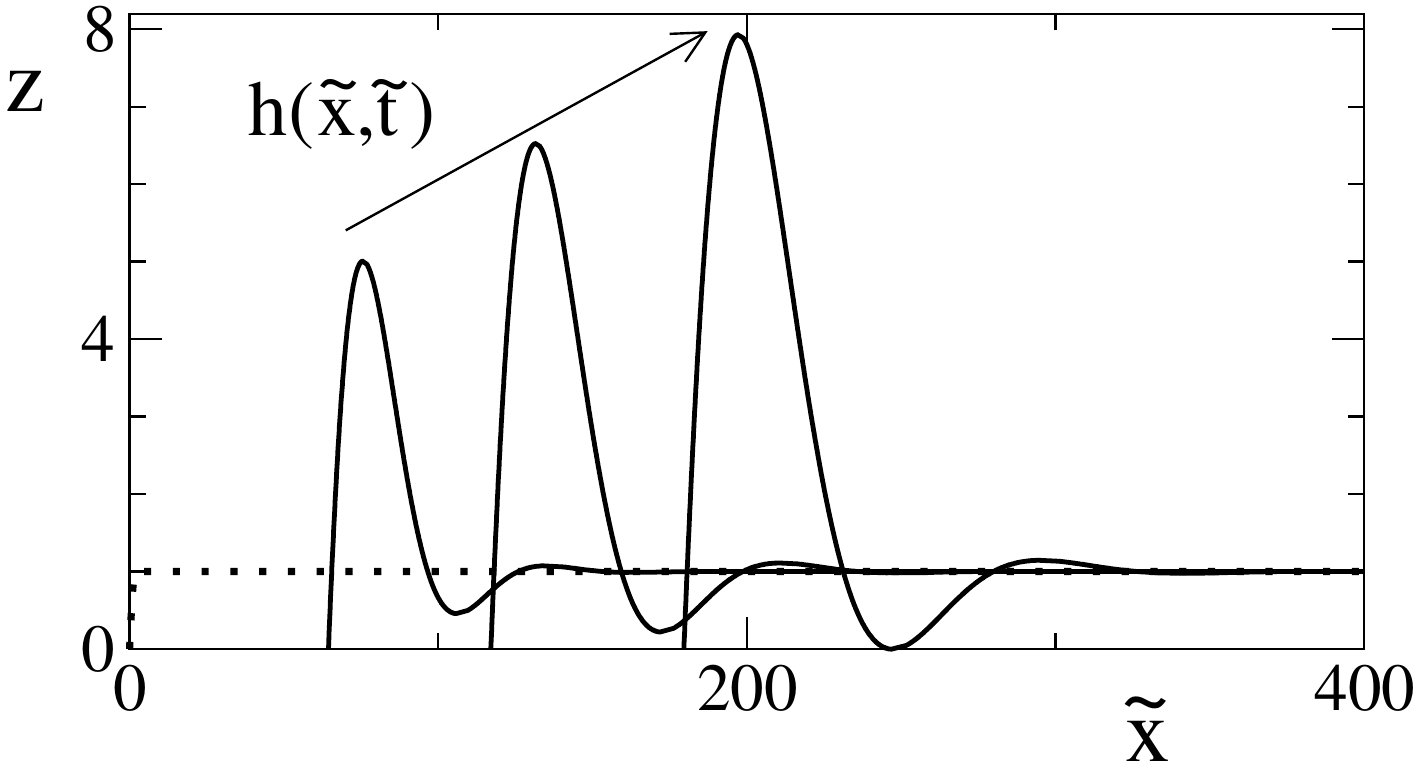}} \\
\subfigure[]{\includegraphics[clip=true,width=0.32\textwidth]{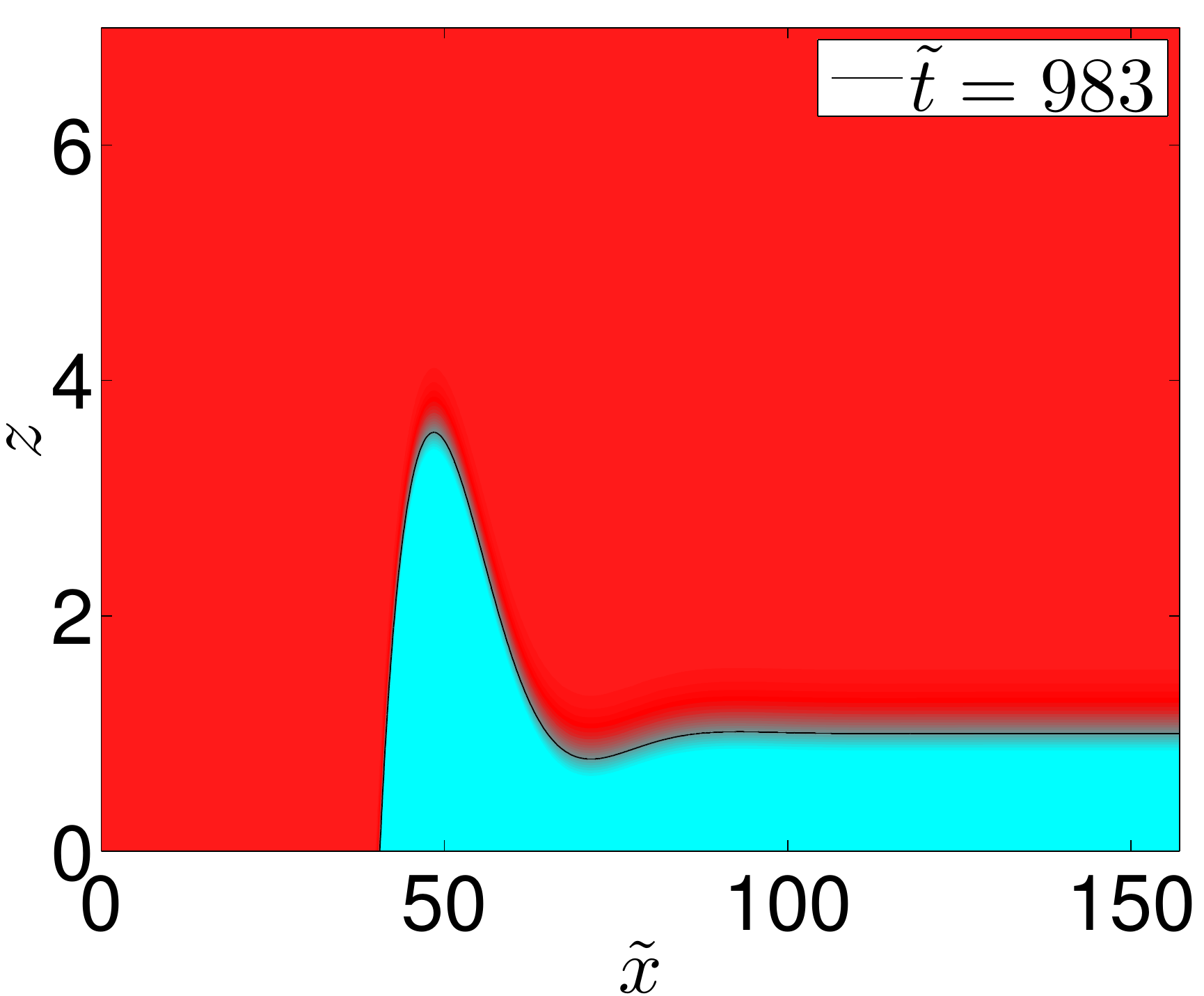}} 
\subfigure[]{\includegraphics[clip=true,width=0.32\textwidth]{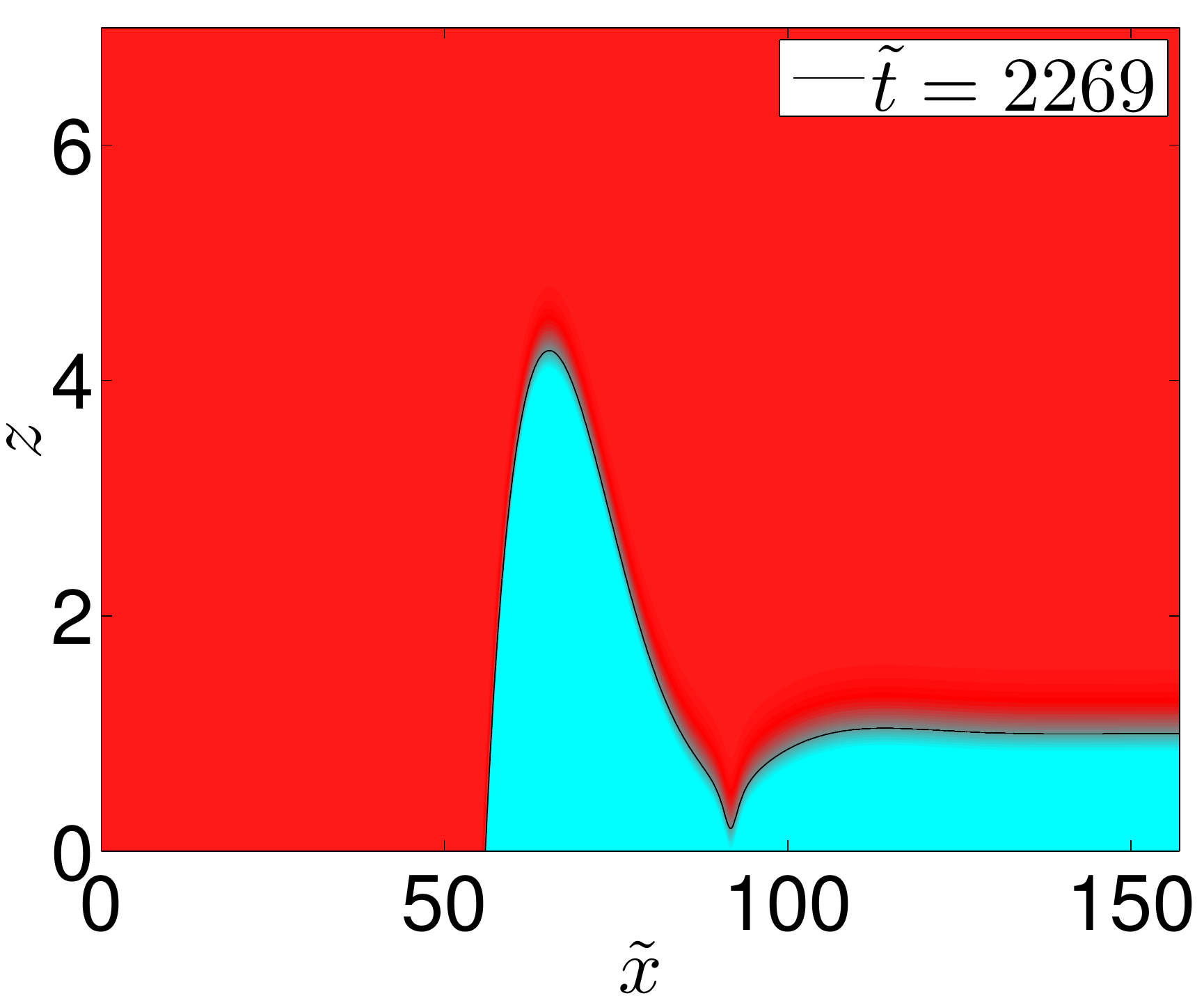}} 
\subfigure[]{\includegraphics[clip=true,width=0.32\textwidth]{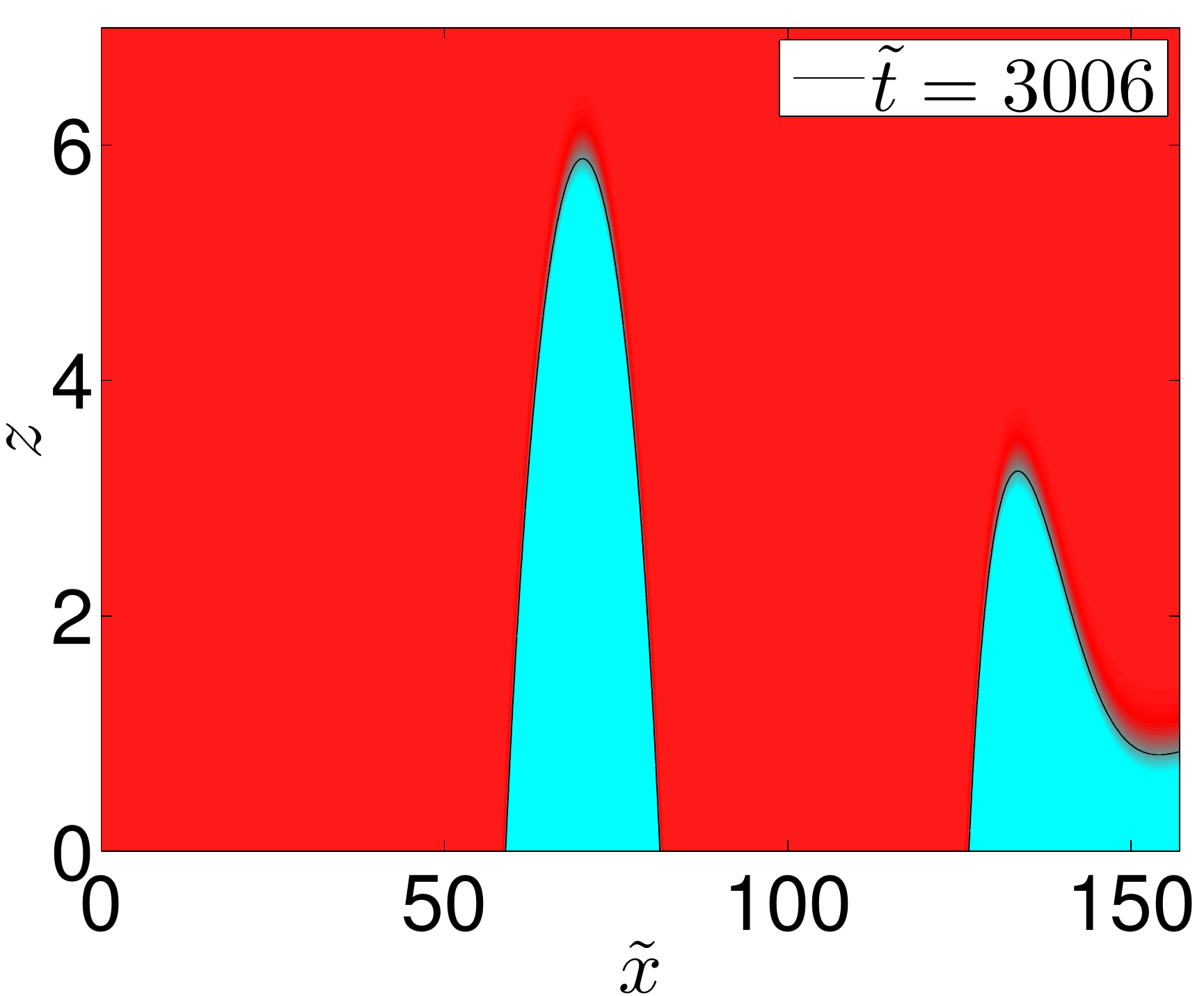}} 
\caption{\label{fig:profs}Evolution of the interface between the two
  phases in the thin-film (top) and phase-field (bottom) models when the initial layers are not of equal
thickness. When this is the case, the location of the second hole depends on the
relative thickness of the layers. In both simulations shown here, the upper layer 
is sufficiently thick
that the second hole is nucleated on the lower substrate, causing a drop to form 
instead of a column. 
Top (a): Simulation of the thin-film model. The initial condition is denoted by a dotted line, and
three profiles are shown at times $\tilde t=2\times 10^3$, $8\times 10^3$
and $2.2\times 10^4$ by solid lines. 
The corresponding contact
lines positions are $\tilde s=64.4$, $117$ and $180$ respectively.
The arrow indicates the direction of time.  The last profile is taken
at the moment where the minimum touches $z=0$, and the height of the ridge at this time is
7.91. 
Bottom (b)--(d): Simulation of the phase-field model with an equilibrium contact angle of $45^\circ$.
When the minimum ahead of the rim is within an $O(\varepsilon)$ distance from the lower substrate,
the rim is rapidly pinched off from the layer which stunts the growth of the ridge. Here, the
height of the ridge is approximately 4.3 when the rim detaches, which is much smaller than 
the corresponding height in the thin-film model.}
\end{figure}

We can use the thin-film model to explore what happens
if the initial horizontal bilayer configuration does not have an A-B ratio that is 50:50,
in which case one of the layers will be thicker than the other. 
If we consider
a receding contact line at the bottom interface, then by setting $d$
to a value larger than two, we have a situation where the top,
i.e., the A-rich layer is thicker.  The receding contact line forms a rim
that eventually hits the top substrate, provided $ d<7.91$. For
larger $d$, the minimum immediately following the rim hits
the bottom substrate first, see Fig.~\ref{fig:profs} (a).  Then, material is
separated from the layer similar to the shedding observed by Wong et
al.~\cite{Wong2000} for surface diffusion.  The material that is left
behind equilibrates into into a droplet configuration that typically
touches only one of the two substrates, rather than both as for the stripes.
The numerical results in Fig.~\ref{fig:profs} (a) were obtained for the
thin-film model \eqref{tfall} with initial condition $\tilde s=0$,
$h(\tilde x,0)= h_i(\tilde x)$, with $h_i$ as
in \eqref{icb}. The channel height was set to $d = 20$, although the
same results would be obtained for any $d > 7.91$ but with different
values of $\tilde{t}$. 
This behaviour is to be expected for thin-film equations with mobility $n<3/2$, as shown e.g. in \cite{KB01,Dziwnik2013}.

The same rim-shedding behaviour can also be observed in the phase-field model;
see Fig.~\ref{fig:profs} (b)--(d).  In this case the thickness ratio of the upper
to lower layer was chosen to be 9:1, and the initial condition was formed using \eqref{icb}
together with \eqref{eqn:si_comp_soln}. The equilibrium contact angle was set to $45^\circ$
with $\varepsilon = 0.32$. The dynamics share some quantitative and qualitative similarities with
the thin-film model. By monitoring the position of the contact line before the rim detaches, we find that it converges to the $t^{2/5}$ behaviour that is predicted from an asymptotic analysis of the thin-film equation \cite{FK04,Dziwnik2013}.
However, a key difference between the models arises when the minimum that follows the rim gets within an
$O(\varepsilon)$ distance from the lower substrate, as the rim in the phase-field model 
is then rapidly pinched off from the main layer (Fig.~\ref{fig:profs} (b)). This causes the rim to detach much
sooner than it does in the thin-film model and as a consequence, the growth of the ridge is
stunted.  In this simulation the height
of the ridge when the rim detaches is approximately 4.3, which is 
significantly smaller than the value of 7.9 found using the thin-film model. 
Upon detaching from
the layer, the rim evolves under the action of interface minimisation to form a droplet with a 
height that is greater than the ridge of the rim; in this case the steady-state height of the droplet
is approximately 6.2. If the thickness of the upper layer is
sufficiently small, the top of the droplet will come into contact with the upper substrate
and a column will form. This suggests that for certain thickness ratios, the transformation
from a bilayer to a sequence of columns occurs via the intermediate processes of rim shedding
and droplet formation. 

\section{Conclusions and Outlook}\label{sec:con}

In this paper, we considered substrate-induced phase separation and the
dynamics of the interfaces for a binary mixture in a thin film geometry,
with a substrate at the top and bottom but unconfined in the lateral direction.
Using a Cahn--Hilliard model that includes appropriate contributions from
the substrate-material interfaces, we explored the conditions under which
multilayer domains form that are separated by thin interface regions.
In particular, we established when the cooling of the mixture below the
critical temperature gives rise to exactly two horizontal layers.

We show that a finite-size perturbation, specifically a hole in one
of the layers, initiates a cascade of transitions into a vertically
striped state.  While this is analysed via direct numerical simulation
of the initial boundary value problem for the Cahn--Hilliard model, we
exploit the multiple length scale separation to derive successively a
sharp-interface model and its corresponding thin-film approximation,
and investigated the validity as an approximation to the original 
phase-field model.

The thin-film model we have derived  belongs to a class of parabolic PDEs for which we can draw on a rich body of literature. For example, it was shown \cite{KB01} that a thin-film model with mobility of $h^n$ with $n<3$ is consistent with a moving contact line and a finite contact angle. Moreover, for dewetting problems, it was shown that for a quadratic mobility,
the static contact angle imposed by the intermolecular potential
also applies to the case where the contact line recedes \cite{MWW06},
and in \cite{FK04} that the microscopic static contact angle is
preserved. Similar arguments can be given for $n=0$, as is the case here, thus lending support to the assumption made here that
the static contact angle
carries over to the dynamical (diffusive) case. 
Our analysis, which is two dimensional, will benefit even more the study of application
relevant three-dimensional counterparts, since now we can exploit our new dimension-reduced thin-film model. A well-known example, symmetry-breaking fingering instability of a receding front in a 3D setting \cite{KB01,ReiteS01,Jiang2003,Dziwnik2013}.

The combination of the phase-field, sharp-interface, and thin-film models developed here provides  efficient descriptions of structure formation as well as long-time dynamics.
For example, if one of the horizontal layers is much thinner than the other, the thin-film model reveals that the cascading rupture events
will only occur in the thinner of the two layers by repeated shedding of the rim, thus leading to an array of droplets
of the minority phase rather than a series of vertical stripes.  While
for antisymmetric substrates the stripes have straight edges, suggesting a very slow
coarsening of the domains, for symmetric 
substrate configurations the stripes are lense-like. It would be interesting to look at the coarsening
behaviour and how it can be captured by the sharp-interface or thin-film model.
If, on the other hand, we start from a structured state with sufficiently narrowly
spaced vertical stripes, a fast coarsening occurs by merging stripes
once initiated by a suitable perturbation. All this demonstrates how the interplay of
geometrical confinement, bulk phase separation, and interface energy effects
give rise to a large variety of structure-forming processes that can be
tuned to achieve design goals for specific technological applications.

\section*{Acknowledgements}
The authors thank Maciek Korzec (Technische Universit\"at Berlin) for his help with the spectral code for the CH equation. VB, AG, MH, AM are grateful for the support by KAUST (Award Number KUK-C1-013-04) and the James Martin School (VB, AG). BW gratefully acknowledges the support by the Federal Ministry of Education (BMBF) and the state government of Berlin (SENBWF) in the framework of the program "Spitzenforschung und Innovation in den Neuen L\"andern" (Grant Number 03IS2151) and the hospitality of OCCAM. AG is a Wolfson/Royal Society Merit Award Holder and acknowledges support from a Reintegration Grant under EC Framework VII.
\bibliographystyle{plain}
\bibliography{paperttb}

\begin{appendix}
\section{Proof of monotonically decreasing free energy}
\label{app:free_energy}
The evolution of the phase-field model given by
\begin{subequations}
\label{app:pf}
\begin{align}
  \phi_t &= \nabla \cdot \left[(1 - \phi^2) \nabla \mu\right], \label{app:e1}\\
  \mu &= \frac{1}{T}\left[f'(\phi) - \varepsilon^2 \Delta \phi\right] \label{app:e2}
\end{align}
with boundary conditions
\begin{alignat}{2}
  \label{app:bc1}
\mu_z&=0, &\quad \eps\mfnd_z&=f_0'((1+\mfnd)/2)\quad \text{at } z=0,\\
\mu_z&=0, &\quad \eps\mfnd_z&=f_0'((1-\mfnd)/2)\quad \text{at } z=d, \label{app:bc2}
\end{alignat}
\end{subequations}
is such that it monotonically decreases the free energy of the system when the temperature
is held at a fixed value. The free energy for this system is given by
\begin{align}
  \frac{F[\mfnd]}{T}  &= \frac{1}{T}\int_{0}^{d}\int_{-\infty}^{+\infty}
  f(\mfnd(x,z)) + 
  \frac{\varepsilon^2}{2} \left|\nabla \mfnd(x,z,t)\right|^2 \,\mathrm{d}x\mathrm{d}z
  \notag
  \\
  &\quad+\frac{2\varepsilon}{T}\int_{-\infty}^{+\infty}
  f_0((1+\mfnd(x,0,t))/2) \,\mathrm{d}x+\frac{2\varepsilon}{T}\int_{+\infty}^{-\infty} f_0((1-\mfnd(x,d,t))/2) \,\mathrm{d}x.
  \notag
\end{align}
To see that this is a monotonically decreasing function of time, we differentiate 
with respect to time and apply the divergence theorem to obtain
\begin{align}
  \begin{split}
  \frac{1}{T}\frac{dF}{dt} &= \frac{1}{T}\int_{0}^{d}\int_{-\infty}^{+\infty}\left[f'(\phi) - \varepsilon^2 \Delta \phi\right]\phi_t\, \mathrm{d}x \mathrm{d}z \\ \quad &+ \frac{\varepsilon}{T} \int_{-\infty}^{+\infty} \left[-\varepsilon \phi_z(x,0) + f_0'((1+\phi(x,0,t))/2)\right]\phi_t(x,0,t)\,\mathrm{d} x \\
  &+ \frac{\varepsilon}{T} \int_{+\infty}^{-\infty} \left[\varepsilon \phi_z(x,d,t) - f_0'((1+\phi(x,d,t))/2)\right]\phi_t(x,d,t)\,\mathrm{d} x.
\end{split}
\end{align}
The boundary integrals vanish due to the boundary conditions in \eqref{app:bc1} and \eqref{app:bc2}, and using 
the bulk equations \eqref{app:e1} and \eqref{app:e2}, this expression can be simplified to
\begin{align}
  \frac{1}{T}\frac{dF}{dt} = \int_{0}^{d}\int_{-\infty}^{+\infty} \mu \nabla\cdot\left[(1 - \phi^2)\nabla \mu\right]\,\mathrm{d}x\mathrm{d}z.
\end{align}
Another application of the divergence theorem yields
\begin{align}
  \frac{1}{T}\frac{dF}{dt} = -\int_{0}^{d}\int_{-\infty}^{+\infty}(1-\phi^2)|\nabla \mu|^2\,\mathrm{d}x\mathrm{d}z,
\end{align}
where the boundary terms vanish because of the no-flux conditions on the substrates. Assuming the
order parameter satisfies $-1 \leq \phi \leq 1$, we have
\begin{align}
  \frac{1}{T}\frac{dF}{dt} \leq 0,
\end{align}
thus completing the proof. 

The same result also holds for the rescaled system in \eqref{beapp2} which has
a free energy given by (after dropping the hats)
\begin{align}
  \begin{split} \label{app:scaled_F}
  F  &= \int_{0}^{d}\int_{-\infty}^{+\infty}
  f(\mfnd(x,z)) + 
  \frac{\varepsilon^2}{2}\left|\nabla \mfnd(x,z,t)\right|^2 \,\mathrm{d}x\mathrm{d}z
  \\
  &\quad+\varepsilon\int_{-\infty}^{+\infty}
  f_0(\phi(x,0,t)) \,\mathrm{d}x+\varepsilon\int_{+\infty}^{-\infty} f_0(-\phi(x,d,t)) \,\mathrm{d}x.
\end{split}
\end{align}

\end{appendix}
\end{document}